\documentclass[12pt]{article}

\usepackage{amsfonts,amssymb,graphics,epsfig,verbatim,bm,latexsym,amsmath,url,amsbsy,natbib,color}

\newtheorem{theorem}{Theorem}
\newtheorem{assumption}{Assumption}

\newtheorem{definition}{Definition}

\def\E{{\mathbb{E}}}
\def\R{{\mathbb{R}}}
\def\P{{\mathbb{P}}}
\def\independenT#1#2{\mathrel{\rlap{$#1#2$}\mkern2mu{#1#2}}}
\newcommand\independent{\protect\mathpalette{\protect\independenT}{\perp}}

\usepackage[left=2.5cm, right=2.5cm, top=2.5cm]{geometry}
\def\spacingset#1{\renewcommand{\baselinestretch}%
{#1}\small\normalsize} \spacingset{1}
\usepackage[colorlinks = true,
            linkcolor = red,
            urlcolor  = blue,
            citecolor = blue,
            anchorcolor = blue]{hyperref}
\begin{document}

 \title{\bf Testing for homogeneous treatment effects in linear and nonparametric instrumental variable models}
\author{
{\large Jad B\textsc{eyhum}}
\footnote{ORSTAT, KU Leuven. Financial support from the European Research Council (2016-2021, Horizon 2020 / ERC grant agreement No.\ 694409) is gratefully acknowledged.}\\\texttt{\small jad.beyhum@gmail.com}
\and
\addtocounter{footnote}{2}
{\large Jean-Pierre F\textsc{lorens}}
\footnote{Toulouse School of Economics, Universit\'e Toulouse Capitole. Jean-Pierre Florens acknowledges funding from the French National Research Agency (ANR) under the Investments for the Future program (Investissements d'Avenir, grant ANR-17-EURE-0010).}
\\\texttt{\small jean-pierre.florens@tse-fr.eu}
\and {\large Elia L\textsc{apenta}}
\footnote{CREST, ENSAE Paris, Institut Polytechnique de Paris}
\\\texttt{\small elia.lapenta@gmail.com}
\and
{\large Ingrid V\textsc{{an} K{eilegom}}\ $^*$}
\\\texttt{\small ingrid.vankeilegom@kuleuven.be}
}

\maketitle

    \begin{abstract}
The hypothesis of homogeneous treatment effects is central to the instrumental variables literature. This assumption signifies that treatment effects are constant across all subjects. It allows to interpret instrumental variable estimates as average treatment effects over the whole population of the study. When this assumption does not hold, the bias of instrumental variable estimators can be larger than that of naive estimators ignoring endogeneity. This paper develops two tests for the assumption of homogeneous treatment effects when the treatment is endogenous and an instrumental variable is available.  The tests leverage a covariable that is (jointly with the error terms) independent of a coordinate of the instrument. This covariate does not need to be exogenous. The first test assumes that the potential outcomes are linear in the regressors and is computationally simple. The second test is nonparametric and relies on Tikhonov regularization. The treatment can be either discrete or continuous. We show that the tests have asymptotically correct level and asymptotic power equal to one against a range of alternatives. Simulations demonstrate that the proposed tests attain excellent finite sample performances. The methodology is also applied to the evaluation of returns to schooling and the effect of price on demand in a fish market.

    \end{abstract}
\vspace{0in}

\spacingset{1.4}
    \section{Introduction}

We are interested in the effect of a (possibly continuous) random treatment $Z$ (with support $\mathcal {Z}\subset\mathbb{R}^{p}$) on a scalar outcome outcome $Y$. Let $Y(z)$ be the scalar potential outcome of the outcome $Y$ under treatment status $z\in\mathcal{Z}$. The potential outcomes are not observed, but $Y=Y(Z)$ is part of the data. We write
$$Y(z)=\varphi(z)+U(z),$$
where $U(z)$ is an error term such that $\E[U(z)]=0$. The goal is to identify the structural function $\varphi$, which allows to obtain the average treatment effect of changing treatment level from $z\in\mathcal{Z}$ to $z'\in\mathcal{Z}$, i.e. $\E[Y(z')-Y(z)]=\varphi(z')-\varphi(z).$

We say that $Z$ is endogenous when the structural function $\varphi$ is not characterized by the distribution of $Y$ conditional on $Z$. Even in this setting, it is possible to identifiy $\varphi$ thanks to an instrumental variable $W$ with support $\mathcal{W}\subset\R^q$. Loosely speaking, an instrumental variable satisfies two conditions. First, it is (in some sense specified in the paper) independent of $\{U(z)\}_z$. Second, it is sufficiently related to $Z$. 

Our goal is to test the following hypothesis
$$H_0: \text{ $U(z) =U$ for all $z\in \mathcal{Z}$}.$$
We call this assumption ``homogeneous treatment effects'' since it stipulates that the treatment effect $Y(z’)-Y(z)$ is a deterministic variable. Under $H_0$, we have $Y=\varphi(Z)+U$, which corresponds to the standard nonparametric instrumental variable (NPIV) model, see \cite{newey2003instrumental,horowitz2007nonparametric,darolles2011nonparametric,chen2012estimation} and the book \cite{li2007nonparametric}, among others.

There are two reasons why one would be interested in testing $H_0$. First, under the instrumental variable conditions, the aforementioned works in the instrumental variable literature show that the structural function $\varphi$ is identified if $H_0$ holds. When $H_0$ is not satisfied, instrumental variable estimators can be more biased than naive estimators ignoring confounding since the error term $U(Z)$ of the model can be dependent of $W$ (through $Z$). In this case, the techniques developed in the nonparametric instrumental variable literature outlined above should not be used. Second, if $H_0$ is true, the treatment effects are constant (i.e. $Y(z’)-Y(z)$ is a deterministic variable). This allows to estimate all counterfactuals for each observation
in the dataset (individualized predictions) using an estimator of $\varphi$. It also means that the treatment does not distort the ranks of potential outcomes, which has implications in terms of inequalities.

The contributions of the paper are as follows. We propose two tests for $H_0$. The tests rely on a scalar covariate $X$ such that $(\{U(z)\}_z,X) $ is independent of one component $W_k$ of the instrument $W$. Importantly, the covariable $X$ is allowed to be dependent of $\{U(z)\}_z$ (even conditional on $W$). The first test relies on the assumption that the potential outcomes are linear in the regressors. In a first step, it estimates $U$ as the residual of a two stage least-squares regression. Then, it tests that the estimator of $U$ is uncorrelated with $X(W_k-\E[W_k])$, which is an implication of homogeneous treatment effects under our assumptions. The second test is nonparametric. In this case, the error term $U$ is estimated by the residual of a Tikhonov regression. The second step of the test is then the same as that of the linear test. In both cases, the treatment can be discrete or continuous. We study the asymptotic distribution of the test statistic and the power of the test. We discuss how to choose $X$: to maximize the likelihood that the conditions on $X$ are satisfied
, we argue in the paper that $X$ should be chosen so as to be independent of $W$. Finally, the empirical performance of the tests is assessed through simulations and illustrated thanks to applications on returns to schooling and price elasticity of demand in a fish market.\\

\textbf{Related literature} Our paper is related to a large literature on testing for (various implications of) homogeneous treatment effects: see \cite{ koenker2002inference, crump2008nonparametric, chernozhukov2005subsampling, ding2016randomization,hsu2017consistent, goldman2018comparing,chung2021permutation, sant2021nonparametric} and \cite{dai2022u} among others. Apart from \cite{sant2021nonparametric}, none of these papers considers the case where there is endogeneity. \cite{sant2021nonparametric} only studies the setting where both the treatment and the instrument are binary and a monotonicity assumption such as in the literature on local average treatment effects holds (see \cite{angrist1996identification}). Instead, our paper allows for confounding and continuous treatments.

Another related literature is that on the hypothesis of rank invariance (see \cite{chernozhukov2005iv}). This assumption is the counterpart of our homogeneous treatment effects condition in nonseparable instrumental variable models (in contrast with the present paper which studies separable models). \cite{chernozhukov2005iv} notes that identification of $\varphi$ can be obtained under a weaker (but less interpretable) form of rank invariance called rank similarity. We could define an assumption analogous to rank similarity in our case of separable models. Our tests would also work for this assumption. We chose to focus solely on homogeneous treatment effects to simplify the exposition. Moreover, \cite{dong2018testing}, \cite{frandsen2018testing} and \cite{kim2022testing} develop tests of the rank invariance assumption in nonseparable instrumental variable models in the case where the treatment and the instrument are binary. The case with continuous treatment is much more challenging because regularization is needed for nonparametric estimation. Note also that \cite{dong2018testing} and \cite{kim2022testing} rely on a monotonicity assumption from local average treatment effects literature (see \cite{angrist1996identification}), while our approach does not require it. The test developed in \cite{frandsen2018testing} bears similarities with our tests since they are based on equivalent single restrictions. However, our models and the first steps of our tests are different. We provide a more in depth analysis of the underlying assumptions of the method and study it in a different context with models for average treatment effects and allow for continuous treatments.

\textbf{Outline} Section \ref{sec.l} presents the linear test along with theory, simulations, and an application to returns to schooling. Then, in Section \ref{sec.np}, we introduce the nonparametric test, discuss its asymptotic properties, evaluate its finite sample performance through numerical experiments, and illustrate the test with an application to demand estimation.  

\section{Linear test}\label{sec.l}\setcounter{equation}{0}

\subsection{Model}\label{subsec.linmodel}
In this section, we develop theory for a test assuming that the mapping $\varphi$ is linear and the instrument $W$ is uncorrelated with $\{U(z)\}_z$, i.e. there exists $\beta\in\R^p$ such that 
$$Y(z)=z^\top\beta+U(z),\ \E[WU(z)]=0,$$
for all $z\in\mathcal{Z}$. 
We also impose the usual relevance assumption for instrumental variables in the linear model, i.e. $\E[WZ^\top]$ has rank equal to $p$. The aim is to recover $\beta$, since the average treatment effect of changing treatment level from $z\in\mathcal{Z}$ to $z'\in\mathcal{Z}$ can be expressed as
$\E[Y(z')-Y(z)]=(z'-z)^\top\beta.$ 

Let the symbol $\independent$ stand for statistical independence. To test for $H_0$, we leverage a scalar covariate $X$ satisfying the following condition:

\begin{assumption}\label{W}
There exists $k\in\{1,\dots, q\}$ such that, for all $z\in\mathcal{Z}$, 
$(\{U(z)\}_z,X)\independent W_k.$
\end{assumption}
This assumption allows $X$ and $\{U(z)\}_z$ to be dependent (even conditional on $W$). Hence $X$ does not need to be exogenous or satisfy an exclusion restriction. In particular, $X$ is not a second instrument.

We argue that Assumption \ref{W} is very likely to hold for some $X$ if $W_k$ is a good instrument. Indeed, if $W_k$ is independent of the unobserved heterogeneity of the model, there is every reason to believe that it is also independent of some observable variables. Note that Assumption \ref{W} is equivalent to the standard instrumental variable condition $\{U(z)\}_z\independent W_k$ and the additional condition $X\independent W_k|\{U(z)\}_z.$ Hence, the condition we add to the standard instrumental variable literature is really $X\independent W_k|\{U(z)\}_z.$

The tests of $H_0$ that we propose depend crucially on Assumption \ref{W}. Contrarily to $H_0$, hypothesis \ref{W} can be heuristically justified. The econometrician should try to select a covariable $X$ which is likely to be (jointly with $\{U(z)\}_z$) independent of $W_k$. In applications, we recommend to pick a variable $X$ independent of $W_k$ since this should make Assumption \ref{W} more likely to hold as $X\independent W_k$ is a necessary condition for Assumption \ref{W}. The independence of $X$ and $W_k$ can be assessed by a statistical test of independence such as $\chi^2$ or Kolmogorov-Smirnov independence tests. Our empirical applications illustrate possible choices of $X$.

Note that Assumption \ref{W} could hold for several components of $W$ and a multidimensional $X$. The extension of the test would be straightforward in this case. We focus on the scalar case to simplify the exposition and because the validity of this assumption for a single component of $W$ is less restrictive.\\

Let us now define the population analog of the two-stage least squares estimator (henceforth, TSLS) in this context:
$$\beta^{TSLS} =\left[\Gamma^\top \E[WW^\top]\Gamma\right]^{-1}\Gamma^\top \E[WY],$$
where $\Gamma =\E[WW^\top]^{-1}\E[WZ^\top].$
In this case, under $H_0$, $\beta$ can be estimated by TSLS, i.e. $\beta^{TSLS}=\beta$. However, when $H_0$ does not hold, then the bias of the TSLS estimator can be larger than that of the ordinary least squares (henceforth, OLS) estimator as illustrated by the following example.\\

\noindent \textbf{Example 1.} We study the case of a randomized experiment with two-sided noncompliance and monotonicity (see \cite{ angrist1996identification}). Let $W=(1,W_2)^\top$, where $W_2$ is a Bernoulli random variable with $\P(W_2=1)=1/2$ and $Z=(1,Z_2)^\top$, with $Z_2= W_21\left\{\epsilon\ge 1/2\right\}+(1-W_2)1\left\{1/2\le \epsilon<3/4\right\}$, where $\epsilon\independent W$ follows a uniform distribution on the interval $[0,1]$. Note that $\P( Z_2=1|W_2=0)=1/4$ and $\P( Z_2=1| W_2=1)=1/2$. For some $\alpha>0$, we also impose $\beta_1=0$ and $\beta_2=\alpha/4$ (i.e. $Y(1,z_2)=(\alpha/4)z_2+ U(Z)$) and $U(1,z_2)=z_2(1\left\{\epsilon\ge 3/4\right\}-(1/4))\alpha$ for $z_2=0,1$). This definition ensures that $E[U(z)]=0,$ for $z=0,1$. This model satisfies all the usual instrumental variable assumptions except $H_0$. The average treatment effect over the whole population is 
$$\Delta =\E[Y(1,1)-Y(1,0)]=\frac{\alpha}{4}.$$
We show in the supplementary material that the population analog of the OLS estimator of $\beta_2$ is given by $$\beta^{OLS}_2= \E[Y|Z_2=1]-\E[Y|Z_2=0]=\frac{\alpha}{3}.$$
Instead, in the context of the present example, it is known (\cite{angrist1996identification}) that $\beta^{TSLS}_2$ is equal to the average treatment effects on the population of compliers. The compliers are the subjects who change treatment status with the instrument, or, equivalently, with $\epsilon\ge 3/4$. As a result, it holds that $\beta_2^{TSLS}=\alpha$.
Hence, the bias of the TSLS estimator is larger than that of the OLS estimator for $\Delta$ and can even go to infinity as $\alpha\to\infty$.
\subsection{Testable implication}\label{sec: testable implication}
Let $U^{TSLS}= Y-Z^\top\beta^{TSLS}$. Under the null hypothesis $H_0$, $U(z)$ will not depend on the treatment level $z$ and $U^{TSLS}=U$, which yields $(U^{TSLS},X)\independent W_k$ (by Assumption 2.1). This last independence condition implies that 
\begin{equation}\label{testable}\E[U^{TSLS}X(W_k-\E[W_k])]=0.
\end{equation}
This is the testable implication of $H_0$ that we use to construct the test. In particular, our test estimates the moment \eqref{testable} and checks if the latter is close to $0$.
Remark that the role of the covariable $X$ appears clearly in \eqref{testable}. Indeed, when $W$ contains an intercept, the equality \begin{equation}\label{testf}\E[U^{TSLS}(W-\E[W])]=0\end{equation} always holds by definition of $U^{TSLS}$, regardless of the validity of $H_0$. Hence, a test based on \eqref{testf} would have no power. The role of $X$ is therefore to give power to our test.

A few remarks are in order. First, there are multiple reasons why we choose to focus on testing the moment condition \eqref{testable} instead of any of the other implications of $H_0$. This is because, under our assumptions, \eqref{testable} is equivalent to saying that the coefficients of the linear projection of $W_k-\E[W_k]$ onto  a constant, $U^{TSLS}$, $X$, and the product $XU^{TSLS}$ are all null. Indeed, the latter coefficients are ``proportional" to $\mathbb{E}[(W_{k}-\E[W_k])(1,U^{TSLS},X,U^{TSLS} X)]$, with 
$\mathbb{E}[(W_{k}-\E[W_k])(1,U^{TSLS})]=0$ by construction and $\mathbb{E}[(W_{k}-\E[W_k])X]=0$ by Assumption \ref{W}, so that the coefficients are null if and only if \eqref{testable} holds (under Assumption \ref{W}). We therefore focus on \eqref{testable} because of this nice interpretation. 

Second, we acknowledge that instead of focusing on \eqref{testable} we could directly test the independence condition $(U^{TSLS},X)\independent W_k$. However, we prefer to focus on \eqref{testable} for three main reasons. First, as we show in the next section, our test based on \eqref{testable} is very simple to implement, and hence it will be attractive to practitioners. Differently,
building a test for the independence condition  $ (U^{TSLS},X)\independent W_k$ would require performing a multi-step estimation, where in a first step we estimate the error $U^{TSLS}$, in a second step we should estimate the cumulative distribution functions of $W_k$ and $(U^{TSLS},X)$, and in a third step we should compare these cumulative distribution functions on the basis of a certain distance. A test of this type would be less attractive to implement in practice. Second, as we show in Section \ref{sec.power}, a test based on the moment condition (\ref{testable}) will have power against a wide range of alternatives. We also remark that focusing on the moment condition in (\ref{testable}) is not necessarily a limitation: while we do not have consistency against all fixed alternatives, we focus the power towards the specific direction of the moment \eqref{testable}. Instead, a test comparing the nonparametric cumulative distribution functions of $(U^{TSLS},X)$ and $W$ would spread the power along all directions, and may have lower power in the direction of our moment \eqref{testable} and, hence, in the direction of the linear projection of $W_k-\E[W_k]$ onto  a constant, $U^{TSLS}$, $X$ and the product $XU^{TSLS}$, as discussed above.

Third, remark that when $X$ is binary (support equal to $\{0,1\}$), \eqref{testable} becomes $\E[U^{TSLS}(W_k-\E[W_k])|X=1]=0$. Hence, in this case our test checks if $U^{TSLS}$ and $W_k$ are uncorrelated (that is $W_k$ is exogenous) in the subpopulations for which $X=0$ and $X=1$.

\subsection{Sample test}
Let us formally outline the test. Consider an i.i.d. sample $\{Y_i,Z_i,W_i,X_i\}_{i=1}^n$ generated from the model of Section \ref{subsec.linmodel}. Let $\widehat{\beta}^{TSLS}$ be the TSLS estimator in this sample, given by 
$$\widehat\beta^{TSLS}=\left[\widehat \Gamma^\top \left(\frac1n\sum_{i=1}^n W_iW_i^\top\right)\widehat \Gamma\right]^{-1}\widehat \Gamma^\top \left(\frac1n\sum_{i=1}^nW_iY_i\right),$$
where $\widehat \Gamma =\left(\frac1n \sum_{i=1}^nW_iW_i^\top\right)^{-1}\frac1n\sum_{i=1}^n W_iZ_i^\top.$ An estimator for $U_i$ is therefore
$$\widehat U_i=Y_i-Z_i^\top \widehat\beta^{TSLS}.$$ 
 Let $\overline{W}_k=n^{-1}\sum_{i=1}^n W_{ki}$.
 The test statistic is then
$$T_n=\frac{1}{\sqrt{n}} \sum_{i=1}^n  \widehat U_iX_i(W_{ki}-\overline{W}_k).$$
We use bootstrap to obtain the p-value of the test, since the variance of the statistic has a tedious expression and variance estimators based on analytical formulas tend to perform poorly in the presence of heteroscedasticity (see e.g. \citet{mackinnon1985some}). The procedure to compute the p-value is as follows.
\begin{enumerate}
    \item Draw $n$ observations with replacement from the sample $\{Y_i,Z_i,W_i,X_i\}_{i=1}^n$ 
    \item Compute the bootstrapped statistic $T_{n,b}^*$ on the bootstrapped sample
    \item Repeat steps 1-2 $B$ times (with $B$ large) so as to get the collection of bootstrapped statistics $\{T_{n,b}^*\,:\,b=1\ldots,B\}$
    \item Compute the {\itshape symmetric} p-value  as 
       $ \frac{1}{B}\sum_{b=1}^B\,1\left\{|T_{n,b}^*-T_n|>|T_n|\right\}\, .$
\end{enumerate}
If the p-value so obtained is smaller than $\alpha$ (the nominal size of the test), the null hypothesis is rejected at the $\alpha$ nominal level. Notice that in the above procedure we are computing a {\itshape symmetric} p-value. An alternative procedure is to compute an {\itshape equal-tailed} p-value, but in our simulations the symmetric one has a satisfying performance.

\subsection{Power analysis}\label{sec.power}
Informally, when $H_0$ does not hold, $U^{TSLS}$ will, in general, not be equal to some $U(z)$ for a fixed $z\in\mathcal{Z}$. Therefore, in this case, there is no reason that $ (U^{TSLS},X)\independent W_k$, and hence, that \eqref{testable}
holds. This gives power to our test. Going further, we argue in this section that it is overwhelmingly likely that \eqref{testable} is wrong when $H_0$ does not hold.
\color{black}
The test will have asymptotic power equal to $1$ under alternative hypotheses for which
\begin{equation}\label{power}\E[U^{TSLS}X(W_k-\E[W_k])]\ne 0.\end{equation}
Since $U^{TSLS}=Y-Z^\top\beta^{TSLS}=U(Z)+Z^\top(\beta-\beta^{TSLS})$, Equation \eqref{power} is equivalent to 
\begin{equation}\label{power2}\E[U(Z)X(W_k-\E[W_k])]+\E[X(W_k-\E[W_k]) Z^\top] (\beta-\beta^{TSLS})\ne 0.\end{equation}
We see that there are two reasons why the test would have power:
\begin{itemize}
\item[(1)] the variable $W_k$ is correlated with $U(Z)X$;
\item[(2)] it holds that $\beta_\ell\ne\beta^{TSLS}_\ell$ and $\E[X(W_k-\E[W_k]) Z^\top]_\ell\ne 0$ for at least one $\ell$ in $\{1,\dots, p\}$.
\end{itemize}

We argue that (1) and (2) are likely to hold in applications. Indeed, statement (1) is probable since $W$ and $Z$ are correlated and $U(Z)$ depends on $Z$. The fact that $E[X(W_k-E[W_k]) Z^\top]\ne 0$ is likely under the assumption that $E[ZW^\top]$ has rank $p$. When $H_0$ does not hold,  $\beta$ and $\beta^{TSLS}$ should be different, they would be equal only under very specific values of $U(Z)$. 

Notice that it would be possible for (1) and (2) to hold  while the test does not have power. This happens when the two terms in \eqref{power2} compensate each other. This case, however, requires very specific data generating processes.

Overall, although there are cases where the test does not have power, the above discussion suggests that the test has power against a wide range of alternative hypotheses. The following example illustrates this claim.\\

\noindent \textbf{Example 2.} Let $(W,E,X)^\top$ be a $3\times 1$ Gaussian vector with mean zero and variance equal to the identity matrix. Let also $Z=W+WE+\rho WX$, $U(Z) = Z(E+\rho X)$, where $\rho \in\R$, and $\beta= 0$ so that $Y=U(Z)=Z(E+\rho X)$. In this case, we have 
\begin{align*}   \E[YW]&= \E[WZ(E+\rho X)]\\
    &=\E[W^2E+W^2E^2+\rho W^2XE+\rho(W^2X+W^2EX+\rho W^2X^2)]=1+\rho^2.
\end{align*}
Moreover, $\E[ZW]=\E[W^2+W^2E+\rho W^2X]=1$. 
In the present context, it is well-known that $\beta^{TSLS}=\E[YW]/\E[ZW]$, which yields $\beta^{TSLS}=1+\rho^2$. 
We have $U^{TSLS} =Y-Z\beta^{TSLS}= Z\left\{(E+\rho X)-(1+\rho^2)\right\}$. Hence, we get 
\begin{align*}
    \E[U^{TSLS}WX]&= \E[(W+WE+\rho WX)\left\{(E+\rho X)-(1+\rho^2)\right\}WX]\\
    &=\rho \E[(WX)^2]-\rho(1+\rho^2)\E[(WX)^2]=-\rho^3.
\end{align*}
As a result, the test has no power only when $X$ and $U$ are independent, i.e. $\rho =0$, which is a degenerate case. \\

In this previous parametric example, we see that the alternatives under which the test has no power have measure equal to 0 for the uniform measure. The example also shows the role of $X$ in providing power to the test, since the latter does not have power only when $X$ and $(Z,\{U(z)\}_z)$ are independent.

\subsection{Asymptotic theory}\label{subsec.ast}
In this section, we state the asymptotic properties of the test statistic. We make the following assumption, which ensures the convergence of the TSLS estimator.
\begin{assumption}\label{TSLS}
$\E[U(Z)^2+X^2+ ||W||_2^2]<\infty$, $\E[ZW^\top]$ exists and has rank $p$ and $\E[WW^\top]$ exists and has full rank.
\end{assumption}
We have the following theorem.

\begin{theorem}\label{AS} Let Assumptions \ref{W} and \ref{TSLS} hold. Then, $T_n$ converges in distribution to a zero mean Gaussian distribution under $H_0$, while $|T_n|/\sqrt{n}\xrightarrow{\P} C\ne 0$ when \eqref{power} holds.
\end{theorem}
The influence function representation of the test statistic is given in Lemma 1.1 in the supplementary material. The asymptotic variance of the test statistic can be derived from this result. Note that the fact that $|T_n|/\sqrt{n}\to C\ne 0$ when \eqref{power} holds implies that the power of the test goes to $1$ as $n$ goes to $\infty$ under alternative hypotheses satisfying \eqref{power}.

\subsection{Simulations} We study the following data generating process. Let $(W_2, E, X, Z_3)$ follow a standard $4$-dimensional Gaussian distribution. We define $Z_2=W_2+E+X$, $Z=(1,Z_2,Z_3)^\top$, $W=(1,W_2,Z_3)^\top$. The variable $Z_2$ suffers from confounding and is instrumented by $W_2$, while the variable $Z_3$ is exogenous. We let $U(z)=(1+\rho z_2)(E+X),$ for all $z\in\{1\}\times \R^2,$
where $\rho\in \R$. The null hypothesis $H_0$ holds when $\rho =0$. The outcome is $Y=U(Z)$ so that the causal regression function of interest is $\varphi= 0$. We set $k=2$, i.e. we use the second component of $W$ to compute the test statistic $T_n=n^{-1/2} \sum_{i=1}^n  \widehat U_iX_i(W_{2i}-\overline{W}_2).$

\begin{figure}[ht]
\centering
\includegraphics[width=9cm,height=9cm]{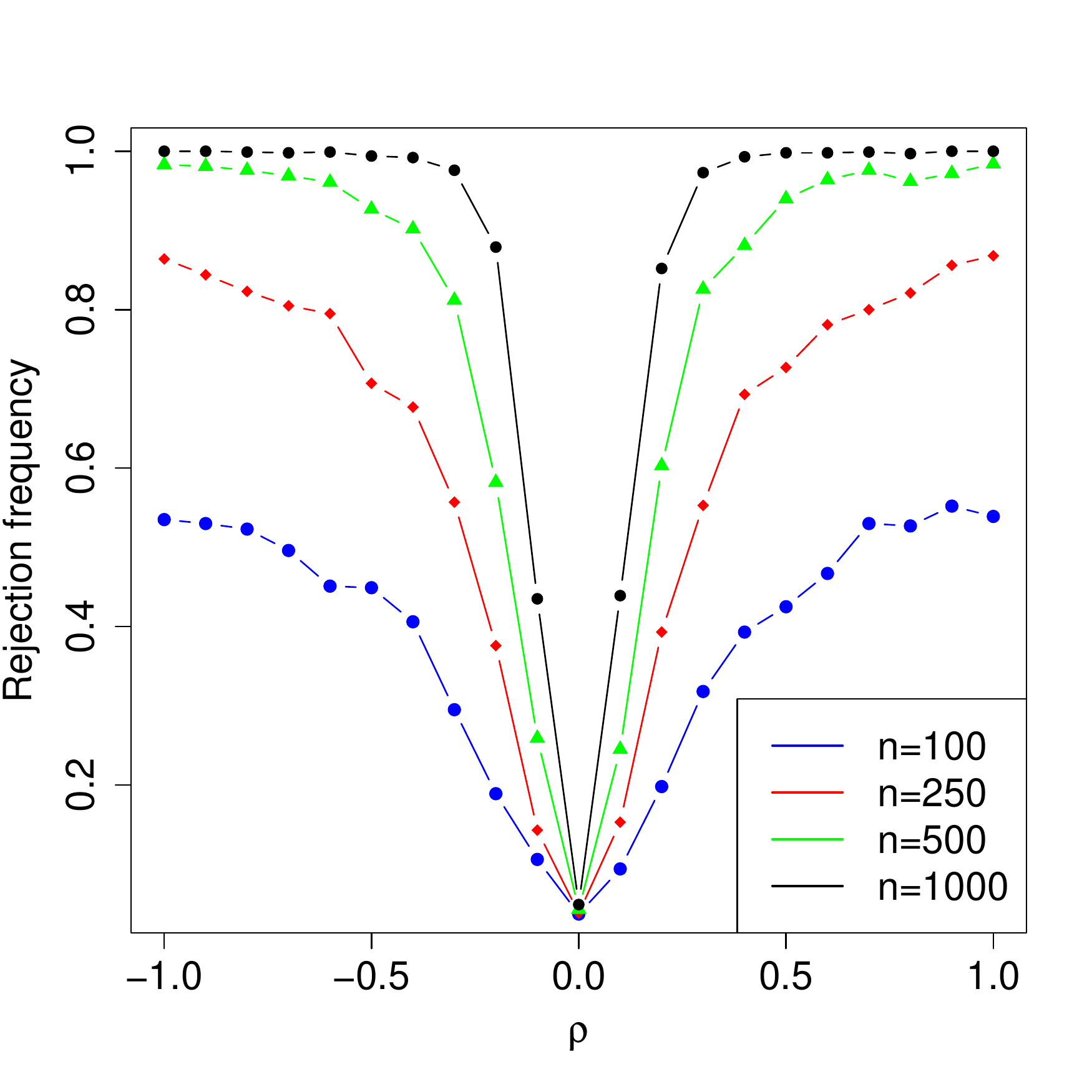}
\caption{Empirical power of the test of theoretical size 5\% as a function of $\rho$.}
\label{linplotpower}
\end{figure}

 We generate data with sample sizes $n\in\{100,200, 500, 1000\}$. We investigate the empirical size of the test when $\rho =0$ (Table \ref{tab:size}).  The results are averages over $10000$ replications using $B=1000$ bootstrap resamples. We also study the empirical power of the test when $\rho=-1, -0.9,\dots,1$, with a thousand replications and bootstrap resamples (See Figure \ref{linplotpower}). The empirical size of the test is almost nominal even for low sample sizes. The power of the test increases as the deviation from the null ($|\rho|$), or the sample size, become larger.

\begin{table}[ht]
\centering
\renewcommand{\arraystretch}{0.8} %
\begin{tabular}{ccccc}
\hline
n & 100 &250 & 500& 1000 \\
\hline
Empirical size at 5\% &0.0397 & 0.0512 & 0.0506& 0.0506\\
Empirical size at 10\% & 0.1038& 0.1050 & 0.1038& 0.1014\\
\hline
\end{tabular}
\caption{Empirical size of the test of theoretical size 5\% and 10\%  for various sample sizes.}
\label{tab:size}
\end{table}

\subsection{Application on returns to schooling}\label{subsec.rtos}
In this section, we test for {homogeneous treatment effects} in a famous application on returns to schooling from \cite{card1993using}. The database is available in the R package \textit{ivreg} and is an extract from the 1976 National Longitudinal Survey (NLS) of young men. Here, $Y=\log(wage)$, $Z=(\boldsymbol{1},education, experience, experience^2, smsa, south, ethnicity)^\top$, where $\boldsymbol{1}$ denotes a constant, {\itshape education} and work $experience$ are measured in years, and $smsa, south, ethnicity$ are controls whose definition can be found in the \textit{ivreg} package. {\itshape Education} and {\itshape experience} are endogenous since they depend on individual’s ability which is unobserved. The instrument is $W=(\boldsymbol{1}, nearcollege, age, age^2, smsa, south, ethnicity)^\top$, where $nearcollege$ is an indicator whose value is equal to $1$ when the individual grew up near an accredited four-year college.  The validity of the instrument is documented in \cite{card1993using}. 

This corresponds to one of the specifications in the original paper of \cite{card1993using}. The variable $X$ is an indicator for being married. It is unlikely that being married is (jointly with the unobserved heterogeneity of the model) dependent from growing up near a four-year college. In fact, a $\chi^2$ test does not reject the hypothesis of independence of $W_2=nearcollege$ and $X=married$ (p-value equal to $0.55$). Hence, we can reasonably assume that $(\{U(z)\}_z,X)\independent W_2$ for all $z\in\mathcal{Z}$, which corresponds to Assumption \ref{W}. Using 10,000 bootstrap resamplings, the p-value of the test is equal to 0.0965, so that the null hypothesis of {homogeneous treatment effects} can be (borderline) rejected at the 10\% level. This result cautions against interpreting the estimates of \cite{card1993using} as causal effects.

\section{Nonparametric test}\label{sec.np}\setcounter{equation}{0}
In this section we extend the linear framework to consider a nonparametric treatment function. We assume that $Z$ and $W$ are unidimensional to simplify the exposition (it is rare to apply nonparametric procedures with multidimensional variables). We continue supposing that Assumption \ref{W} holds and we keep the notation previously introduced. We write 
$$Y(z)=\varphi(z)+U(z),\  \E[U(z)|W]=0, \text{ for all }z\in\mathcal{Z}\, ,$$
where the conditional mean independence $\mathbb{E}[U(z)|W]=0$ is a direct consequence of Assumption \ref{W}.
The functional form of $\varphi$ is unknown.  
Let $L^2(Z)$ be the set of functions that are square integrable with respect to the distribution of $Z$. We maintain the following assumption which is also standard in the NPIV literature.
 \begin{assumption}\label{completeness} (a) For all $\phi\in L^2(Z)$, $\mathbb{E}[\phi(Z)|W]=0\Rightarrow \phi\equiv 0$. (b) There exists at least one $\widetilde \varphi\in L^2(Z)$ satisfying $\mathbb{E}[Y|W]=\mathbb{E}[\widetilde \varphi(Z)|W]$.
 \end{assumption}
 The first part of the assumption is the completeness condition standard in the NPIV literature. It means that $W$ is sufficiently associated with $Z$. Under the null hypothesis $H_0$, such a condition is necessary and sufficient for identifying $\varphi$  (see, among others, \cite{carrasco_chapter_2007}, \cite{darolles2011nonparametric}, \cite{newey_instrumental_2003}, \cite{dhaultfoeuille_completeness_2011}). The second part states that the NPIV model in the above equation ($\mathbb{E}[Y|W]=\mathbb{E}[\widetilde \varphi(Z)|W]$) is well specified. We will discuss this assumption in Section \ref{subsec: Assumptions and asymptotic behavior}. \\

Recall that the hypothesis that we want to test is 
$$H_0: \text{ $U(z) =U$ for all $z\in \mathcal{Z}$}.$$
Under the null hypothesis of {homogeneous treatment effects} $\mathbb{E}[U(Z)|W]=\mathbb{E}[U(z)|W]=\E[U(z)]=0$, so  the treatment function $\varphi$ must satisfy the following (functional) equation in $\phi\in L^2(Z)$
\begin{equation}\label{eq: integral equation}
\mathbb{E}[Y|W]=\mathbb{E}[\phi(Z)|W]\, .    
\end{equation}

Under the completeness condition (Assumption \ref{completeness} (a)) the solution to the above equation is {\itshape unique} and can be consistently estimated (we introduce an estimator below). Since under the null hypothesis of {homogeneous treatment effects} the treatment function $\varphi$ will satisfy the above equation, $\varphi$ will be identified as its unique solution. Thus, under the null hypothesis of {homogeneous treatment effects} we can consistently estimate the nonparametric treatment function $\varphi$. Differently, when $H_0$ does not hold, the solution to Equation (\ref{eq: integral equation}) will be different, in general, from the treatment function of interest $\varphi$. This solution, denoted with $\widetilde{\varphi}$, is introduced in Assumption \ref{completeness} (b). So, if $H_0$ is not true, although we will be able to estimate consistently the solution to Equation (\ref{eq: integral equation}), we will not in general obtain a consistent estimator of $\varphi$.  It is therefore crucial to test for the {homogeneous treatment effects} hypothesis. 

Let $\widetilde{U}(Z):=Y(Z)-\widetilde{\varphi}(Z)$ and $\widetilde{\varphi}$ denote the unique solution to Equation \eqref{eq: integral equation} (which exists by Assumption \ref{completeness} (ii)). Under the null of homogeneous treatment effects $\widetilde{U}(Z)=U$ and $(X,\widetilde{U}(Z))\independent W$. Thus, by arguing similarly as in the previous section, we test
\begin{equation}\label{eq: single moment for NPIV}
\mathbb{E}[\widetilde{U}(Z)X(W-\mathbb{E}[W])]=0\,
\end{equation}
Remark that, in the NPIV model, instead of conducting a test based on the moment condition in (\ref{eq: single moment for NPIV}), we could build a test to directly check that $(\widetilde{U}(Z),X)\independent W$. However, we choose to focus on the single moment condition in (\ref{eq: single moment for NPIV}) for several reasons. First, our test based on a single moment restriction has power under a wide range of alternatives. Indeed, a power analysis similar to that in Section \ref{sec.power} could be carried out in the present nonparametric case. Some examples of alternative hypotheses under which the test has power are given in Section S2.2 of the supplementary material. Second, as argued in Section \ref{sec.power}, our test avoids to spread power across multiple directions. Third, focusing on a single restriction allows to propose a test which is simple to implement. In contrast, building a test for the condition $(\widetilde{U}(Z),X)\independent W$ would be theoretically intricate. In fact, it would require to estimate the error $\widetilde{U}(Z)$ in a first step and then to estimate the cumulative distribution functions of $W$ and $(\widetilde{U}(Z),W)$ in a second step. It is likely that in this case we would need an $n^{-1/4}$ convergence rate for the first-step NPIV estimator of $\widetilde{\varphi}$, as it is required  in multi-step estimations with a nonlinear second step, see \citet{CLV}. This would require a slow convergence rate of the regularization parameter to allow the variance of the first-step NPIV estimator to converge fast enough to zero, see Assumption \ref{bandwidth} ahead. At the same time, we would need the regularization bias of $\widetilde \varphi$ to disappear fast enough, and this would require that the regularization parameter converges quickly to zero, see Assumption \ref{bandwidth} ahead. Therefore, there would be some tensions between these two conditions and the construction of the test would become intricate. In particular, the rates of convergences obtained with the Tikhonov regularization scheme would not allow to obtain a consistent nonparametric test. A more complicated first-step estimator would be required.

Let $\widehat{\varphi}$ be an estimator for $\varphi$ (below we define our estimator), and let us define the estimated residuals $\widehat{U}_i=Y_i-\widehat{\varphi}(Z_i)$, for $i=1,\ldots,n$.
Our statistic is the empirical counterpart of the above moment  
\begin{equation*}
    S_n=\frac{1}{\sqrt{n}}\,\sum_{i=1}^n \widehat{U}_i\,(W_{i}-\overline{W})\,X_i\, ,
\end{equation*}
where $\overline{W}=n^{-1}\sum_{i=1}^nW_i$ is the empirical mean of $W$. 
In the following section, we explain how to compute the above statistic and implement the test based on it. In Section \ref{subsec: Assumptions and asymptotic behavior}, we present the assumptions and the results for the validity of the test. Next, in Section \ref{subsec:simulations}, we provide evidence about the finite sample performance of the test based on $S_n$. Finally, we illustrate the procedure by an application to demand estimation in Section \ref{sec.emp}.

\subsection{Construction of the statistic and implementation of the test }\label{sec: implementation of NPIV statistic}
In this section, we outline the computation of the statistic $S_n$ and implement the test. To this end, we need (a) to compute $\widehat{\varphi}(Z_i)$ for $i=1,\ldots,n$, and (b) to obtain the p-value necessary for testing. \\

\noindent \textbf{Computation of $\widehat{\varphi}$}. 
Let $\pi$ and $\tau$ be positive functions on $\mathbb{R}$. We denote with $L^2_\pi(\mathbb{R})$ the space of square integrable functions with respect to $\pi$, and let $L^{2}_\tau(\R)$ be similarly defined. The functions $\pi$ and $\tau$ are introduced for the aim of generality and for technical reasons. In simulations they are set to Gaussian density functions. Let us briefly explain their roles. First, from a technical point of view it would be ideal to work with the spaces $L^2(Z)$ and $L^2(W)$, i.e. the spaces of square integrable functions with respect to the distribution of $Z$ and $W$. However, since we do not know such distributions, it is convenient to replace $L^2(Z)$ and $L^2(W)$ with $L^2_{\tau}(\mathbb{R})$ and $L^2_{\pi}(\mathbb{R})$ and work with the latter spaces. Second, by working with $L^2_{\tau}(\mathbb{R})$ and $L^2_{\pi}(\mathbb{R})$ we can obtain that the estimator of the dual of $A$ (defined below) is actually the dual of the estimator of $A$. This is an important property for establishing the asymptotic normality of our statistic $S_n$ under $H_0$. 

Let $f_W$ stand for the density of $W$ with respect to the Lebesgue measure, and assume that $\tau(w)>0$ whenever $f_W(w)>0$. Then, Equation (\ref{eq: integral equation}) is equivalent to

\begin{equation}\label{eq: integral equation 2}
\mathbb{E}[Y|W]\frac{f_W(W)}{\tau(W)}=\mathbb{E}[\widetilde{\varphi}(Z)|W]\frac{f_W(W)}{\tau(W)}.    
\end{equation}

To build an estimator for $\widetilde{\varphi}$ we start from the above integral equation.
Let us denote by $f_{Z W}$ the joint density of $(Z,W)$ with respect to the Lebesgue measure and introduce $L^2_{\pi \otimes \tau}(\R^{2})$, which is the set of functions from $\R^2$ to $\R$ square integrable with respect to the product measure $\pi \otimes \tau$ (the measure generated by the product density $\pi \cdot \tau$). By assuming that $f_{Z W}/(\pi\,\tau)\in L^2_{\pi \otimes \tau}(\R^{2})$, we define
$A: L^2_\pi(\R)\mapsto L^2_\tau(\R)$ as the operator 
$$(A\phi)(\cdot)=\int_{ }\phi(z) f_{ZW}(z,\cdot)dz\,\frac{1}{\tau(\cdot)}.$$
Its Hilbert adjoint  $A^*:L^2_\tau(\R)\mapsto L^2_\pi(\R)$ is given by
$$(A^*\psi)(\cdot)=\int_{ }\psi(w)f_{Z W}(\cdot,w)\,dw\,\frac{1}{\pi(\cdot)}.$$
We also define the “left hand side" of the integral equation (\ref{eq: integral equation 2}) as
$$r(\cdot)=\mathbb{E}[Y|W=\cdot]\frac{f_W(\cdot)}{\tau(\cdot)}.$$
We estimate $A$ and $A^*$ by replacing $f_{ZW}$ with its kernel estimator 
$$\widehat{f}_{ZW}(z,w) =\frac{1}{n\, h_Z\,h_W}\sum_{i=1}^n K_Z\left(\frac{Z_i-z}{h_Z}\right)K_W\left(\frac{W_i-w}{h_W}\right),$$
where $K_Z$ and $K_W$ are two kernel functions while $(h_Z,h_W)$ denote two bandwidths converging to zero as the sample size increases. The mapping $r$ is instead estimated by 
\begin{equation*}
   \widehat{r}(\cdot):= \frac{1}{n h_W}\sum_{i=1}^n Y_i K_W\left( \frac{W_i-\cdot}{h_W} \right)\frac{1}{\tau(\cdot)}.
\end{equation*}
To estimate $\widetilde{\varphi}$, i.e. the solution to Equation (\ref{eq: integral equation 2}), we use a Tikhonov regularization scheme  
$$\widehat \varphi =\left(\lambda I+\widehat A^*\widehat A\right)^{-1}\widehat A^* \widehat r,$$
where $I$ stands for the identity operator, while $\lambda>0$ denotes the Tikhonov regularization parameter. See, e.g., \cite{carrasco_chapter_2007} and \cite{darolles2011nonparametric}.

To compute $\widehat{\varphi}$ we need to select two bandwidths $(h_Z,h_W)$ and a regularization parameter $\lambda$. We will explain below how to do it. For the moment, let us outline the computation of $\widehat{\varphi}$ in the previous display for given $(h_Z,h_W,\lambda)$.
Although $\widehat{\varphi}$ seems to have an abstract expression, its computation is straightforward as it reduces to matrix products. To see this, we first  approximate $\widehat{A}$ and $\widehat{A}^*$ by using a common bias computation similarly as in \citet{centorrino_additive_2017} :
\begin{align*}
    (\widehat{A}\phi)(w)=&\frac{1}{n h_Zh_W}\sum_{i=1}^n K\left( \frac{W_i-w}{h_W} \right)\,\frac{1}{\tau(w)}\,\underbrace{\int_{ }K\left( \frac{Z_i-z}{h_Z}\right)\,\phi(z)\,d z}_{\approx h_Z\,\phi(Z_i)}\\
    \approx & \frac{1}{n h_W}\sum_{i=1}^n K\left( \frac{W_i-w}{h_W} \right)\,\frac{1}{\tau(w)}\,\phi(Z_i)\\
    (\widehat{A}^*\psi)(z)&= \frac{1}{n h_Z h_W}\sum_{j=1}^n K\left(\frac{Z_j-z}{h_Z}\right)\,\frac{1}{\pi(z)}\,\underbrace{\int_{ }K\left( \frac{W_j-w}{h_W}\right)\,\psi(w)\, d w}_{\approx h_W\,\psi(W_j)} \\
    \approx & \frac{1}{n h_Z}\sum_{j=1}^n K\left(\frac{Z_j-z}{h_Z}\right)\,\frac{1}{\pi(z)}\psi(W_j)\, .
\end{align*}

The above approximations hold for $h_Z\, ,\, h_W\, \rightarrow 0$. Let $M_Z$ be the $n\times n$ matrix  having on the $i$th row and $j$th column the element $K((Z_i-Z_j)/h_Z)/[\pi(Z_i) n h_Z]$ and let $M_W$ be the $n\times n$ matrix  having on the $i$th row and $j$th column the element $K((W_i-W_j)/h_W)/[\tau(W_i) n h_W]$. Let $\overrightarrow{\widehat \varphi}:=(\widehat{\varphi}(Z_1),\ldots ,\widehat{\varphi}(Z_n))^\top$. For a generic function $\psi$ of $W$ let $\overrightarrow{\widehat{A}^*\psi}:=((\widehat{A}^*\psi)(W_1),\ldots ,(\widehat{A}^*\psi)(W_n))^\top$. Let $\overrightarrow{\widehat r}$ and $\overrightarrow{\widehat A \phi}$ (for a generic function $\phi$) be similarly defined. By the expression of $\widehat{\varphi}$ and the above approximations we get 
\begin{equation*}
    \overrightarrow{\widehat{r}}=\lambda \vec{\widehat \varphi}+\widehat{A}^*(\widehat{A}\widehat{\varphi})\approx \lambda \overrightarrow{\widehat \varphi} + M_Z \overrightarrow{\left(\widehat{A}\widehat{\varphi}\right)} \approx \lambda \overrightarrow{\widehat \varphi} + M_Z M_W \overrightarrow{\widehat \varphi}  \, ,
\end{equation*}
so that
\begin{equation}\label{eq: implementation of phihat}
    \overrightarrow{\widehat{\varphi}}\approx\left(\,\lambda I + M_Z\,M_W\,\right)^{-1} M_Z\,\overrightarrow{\widehat{r}}\, 
\end{equation}
up to an approximation error. Thus, the computation of $\widehat{\varphi}$ at the data points reduces to a simple matrix computation and the residuals can be easily computed as  $\widehat{U}_i=Y_i-\widehat  \varphi(Z_i)$ with $i=1,\ldots,n$. 

Let us turn now to the selection of the smoothing parameters $(h_Z,h_W)$ and  the regularization parameter $\lambda$. The bandwidths $h_Z$ and $h_W$ are selected by the Silverman Rule of Thumb, so that $h_Z=\widehat{\sigma}_Z\,n^{-1/5}$ and $h_W=\widehat{\sigma}_W\,n^{-1/5}$, where $\widehat{\sigma}_Z$ and $\widehat{\sigma}_W$ denote the sample standard deviations of $Z$ and $W$.  Finally, to  select the Tikhonov regularization parameter, we use the Cross-Validation method developed in \citet{centorrino_additive_2017}. Thus, 
\begin{equation}\label{eq: cross-validation}
    \widehat{\lambda}:=\arg \min_{\lambda} \sum_{i=1}^n \left[(\widehat{A}\widehat{\varphi}^\lambda)_{-i}(W_i)-\widehat{r}(W_i)\right]^2\, ,
\end{equation}
where $(\widehat A \widehat{\varphi}^{\lambda})_{-i}(W_i):=\left[ \widehat{A}_{-i}(\lambda I + \widehat{A}^*\widehat{A})^{-1}\widehat{A}^*_{-i}\widehat{r}\right](W_i)$. Here  $\widehat{A}_{-i}$ and  $\widehat{A}^*_{-i}$ denote the “leave-one-out" versions of $\widehat{A}$ and $\widehat{A}^*$ that use the entire sample except the $i$th observation. By the approximations of $\widehat{A}$ and $\widehat{A}^*$ previously outlined, the vector $\{(\widehat A \widehat{\varphi}^{\lambda})_{-i}(W_i):\ i=1,\ldots,n\}$ is computed as 
\begin{align*}
    \left[(\widehat A \widehat{\varphi}^{\lambda})_{-i}(W_i)\right] _{i=1,\ldots,n}=&\left[ \widehat{A}_{-i}(\lambda I + \widehat{A}^*\widehat{A})^{-1}\widehat{A}^*_{-i}\widehat{r}\right](W_i)_{i=1,\ldots,n} \\
    \approx &\left(M_W-diag(M_W)\right)\,\left(\lambda I + M_Z\,M_W\right)^{-1}\left(M_Z-diag(M_Z)\right)\, \overrightarrow{\widehat r}\, ,
\end{align*}
where $diag(M_W)$ denotes the diagonal matrix having the same main diagonal as $M_W$, and $diag(M_Z)$ is similarly defined. The objective function in (\ref{eq: cross-validation}) has a U-shaped form as a function of $\lambda$ and can be minimized in a simple way by numerical methods, see \citet{centorrino_additive_2017}.

For the sake of clarity, we sum up the steps needed for computing $S_n$ as follows:
\begin{enumerate}
    \item Select $\widehat{\lambda}$ according to the Cross-validation method in Equation (\ref{eq: cross-validation})
    \item Use $\widehat{\lambda}$ to compute $\{\widehat{\varphi}(Z_i):i=1,\ldots,n\}$  as in (\ref{eq: implementation of phihat}) and compute the residuals as $\{\widehat{U}_i=Y_i-\widehat{\varphi}(Z_i):i=1\ldots,n\}$ 
    \item Compute the statistic as $S_n=(1/\sqrt{n})\sum_{i=1}^n\widehat{U}_i(W_{i}-\overline{W})X_i$.
\end{enumerate}

\noindent \textbf{Implementation of the test}.  To implement the test, we are just left with the computation of the p-value. As we show in the next section, under the null hypothesis the statistic $S_n$ is asymptotically normal, but the asymptotic covariance has an intricate expression. So, to obtain the p-value necessary for testing we rely on the pairwise bootstrap. The validity of the pairwise bootstrap is confirmed by our simulations and is not surprising given that we show asymptotic normality. In fact, under different conditions, \cite{CLV} show the validity of the bootstrap for a {\itshape real-valued} estimator based on a first step {\itshape infinite-dimensional} estimator, as is $\widehat\varphi$ in this paper. The procedure goes as follows:
\begin{enumerate}
    \item Draw $n$ observations with replacement from the sample $\{Y_i,Z_i,W_i,X_i\}_{i=1}^n$ 
    \item Compute the bootstrapped statistic $S_{n,b}^*$ on the bootstrapped sample using the same bandwidths $(h_Z,h_W)$ and regularization parameter $\lambda$ as in the original sample
    \item Repeat steps 1-2 $B$ times (with $B$ large) to obtain the collection of bootstrapped statistics $\{S_{n,b}^*\,:\,b=1\ldots,B\}$
    \item Compute the {\itshape symmetric} p-value  as 
       $ \frac{1}{B}\sum_{b=1}^B\,1\left\{|S_{n,b}^*-S_n|>|S_n|\right\}\, .$
\end{enumerate}

\subsection{Asymptotic behavior}\label{subsec: Assumptions and asymptotic behavior}
In this section, we outline the assumptions under which we obtain the asymptotic properties of the test statistics. The following definition introduces some regularity features that the joint density $f_{ZW}$ and the NPIV function $\widetilde{\varphi}$ must satisfy. \\

\begin{definition} For a given function $\gamma$ and for $\alpha \geq 0\, , s > 0$, the space $\mathcal{B}^{s,\alpha}_\gamma (\mathbb{R}^{\ell})$ is the class
of functions
 $g : \mathbb{R}^{\ell} \mapsto \mathbb{R}$ 
 satisfying: $g$ is everywhere $(m-1)$ -times partially differentiable for
$m-1<s\leq m$; for some $R>0$ and for all $x$, the inequality
\begin{equation*}
    \sup_{y:\|y-x\|<R}\frac{\left|g(y)-g(x)-Q(y-x)\right|}{\|y-x\|^s}\leq \psi(x)
\end{equation*}
holds true, where $Q=0$ when $m=1$ and $Q$ is an $(m-1)$th degree homogeneous polynomial in $(y-x)$ with coefficients the partial derivatives of $g$ at $x$ of orders 1 through $m-1$ when $m>1$; $\psi$ is uniformly bounded by a constant when $\alpha=0$ and the functions $g$ and $\psi$ have finite $\alpha$th moments with respect to $1/ \gamma$ when $\alpha>0$, i.e. $\int |g(x)|^\alpha / \gamma(x)\, dx < \infty$ and  $\int |\psi(x)|^\alpha / \gamma(x)\, dx < \infty$\, .
\end{definition}

Let $f_Z$ denote the density of $Z$ with respect to the Lebesgue measure and let us define
$$g(z):=\mathbb{E}[(W-\mathbb{E}[W]) X|Z=z]f_Z(z)\, .$$

\begin{assumption}\label{ differentiabilty and integrability}
$\widetilde{\varphi}\in \mathcal{B}^{\rho,0}_1(\R)\cap L^2_{\pi}(\R)$ and $f_{Z W}/(\pi\,\tau) \in  \mathcal{B}^{\rho,0}_1(\R^{2})\cap L^2_{\pi\otimes \tau}(\R^{2})$ for a $\rho$ specified below, $\mathbb{E}[U^2|W=\cdot]\,f_W/\tau $ is bounded,   $\mathbb{E}[X|Z=\cdot]f_Z/\pi \in L^2_\pi (\R)$, $\mathbb{E}[X^2(W-\mathbb{E}[W])^2|Z=\cdot]f_Z/\pi$ is bounded, and $\mathbb{E}[X^2(W-\mathbb{E}[W])^2|Z=\cdot]$ is bounded.
\end{assumption}

Assumption \ref{ differentiabilty and integrability} is a common regularity condition allowing for several degrees of integrability and differentiability \citep{florens_instrumental_2012}. Under the above assumption $\widehat{A}:L^2_\pi (\R)\mapsto L^2_\tau (\R)$, $\widehat{A}^*:  L^2_\tau (\R) \mapsto L^2_\pi (\R)$, and $\widehat{r}\in L^2_\tau (\R)$. Notice that $\widehat{A}^*$ is actually the Hilbert adjoint of $\widehat{A}$. This aspect is used multiple times in the proofs. 

\begin{assumption}\label{kernel}
$K_Z$ and $K_W$ are symmetric kernels of order $\rho$ with bounded support.  
\end{assumption}
The kernel orders in Assumption \ref{kernel} are assumed to be equal only for notational simplicity. To simplify our theoretical exposition, we further assume that $h_W=h_Z$ and denote each bandwidth by $h$. However, our proofs also hold when such smoothing parameters are set to different values (and the kernels have different orders).

The order $\rho$ of the kernels $K_Z$ and $K_W$, the bandwidth $h$ and the regularization parameter $\lambda$ satisfy the following assumption.
\begin{assumption}\label{bandwidth}
$n h^{2}\lambda^{3/2} \rightarrow \infty$, $n\lambda^2\rightarrow 0$, $h^\rho \lambda^{-3/4}\rightarrow 0$.
\end{assumption}
Such conditions allow us to control the 
regularization bias and the variance of $\widehat{\varphi}$ appearing in the expansion of $S_n$. In particular, the condition on  $n h^2 \lambda^{3/2}$ allows controlling the "variance term". The conditions on $n \lambda^2$ and $h^\rho \lambda^{-3/4}$  
allow us to show that the regularization bias that appears in the expansion of $S_n$ is negligible. This bias is 
due to the ill-posed nature of the inverse problem.

Let us denote with $\left<\cdot,\cdot \right>$ the inner product of either $L^2_\pi(\mathbb{R})$ or $L^2_\tau(\mathbb{R})$, and let $\|\cdot\|$ be the norm induced by such an inner product. The specific inner product or norm we refer to will be clear at each time from the context. Let $(\lambda_j,\varphi_j,\psi_j)_j$ be the Singular Value decomposition of the operator $A$, where $(\lambda_j)_j$ is the sequence of singular values in $\mathbb{R}$, $(\varphi_j)_j$ is an orthonormal sequence in $L^2_\pi(\mathbb{R})$, and $(\psi_j)_j$ is an orthonormal sequence in $L^2_\tau(\mathbb{R})$ (see \cite{kress_linear_1999}). The following assumptions introduces the usual source conditions. 
\begin{assumption}\label{source condition} 
Let $\left<\cdot,\cdot\right>$ denote the inner product on $L^2_{\pi}(\R)$. (a) For some $\eta \geq 2$, $\sum_j \frac{|\left< g, \varphi_j \right>|^2}{\lambda_j^{2\eta}}<\infty $; (b) For some $\theta\geq 2$,
$\sum_j\frac{|\left< \widetilde{\varphi}, \varphi_j \right>|^2}{\lambda_j^{2\theta }}<\infty$.
\end{assumption}

Source conditions are standard in the NPIV literature (see \cite{carrasco_chapter_2007}  or  \cite{darolles2011nonparametric}). In general, the source condition is imposed only on $\widetilde{\varphi}$ when the interest is on estimating $\widetilde{\varphi}$. Here, we need a source condition on both $\widetilde{\varphi}$  and $g$ to establish the asymptotic properties of the statistic based on the residuals from the nonparametric regression. To this end, we also need some regularity conditions on the estimator $\widehat{\varphi}$. So, let  $N_{[\cdot]}(\epsilon,\Phi,||\cdot||_\P)$ be the bracketing  number of size $\epsilon$ of a class of functions $\Phi$, where $||\cdot||_\P$ denotes the $L_2$ norm with respect to the probability law of $Z$, see Van der Vaart (1998). 
\begin{assumption}\label{Donsker} There exists a class of functions $\Phi$ such that
$\int_{0}^1 \sqrt{\log N_{[\cdot]}(\epsilon,\Phi,||\cdot||_\P)} d\epsilon$ $ < \infty $
 and $\P(\widehat{\varphi}\in \Phi )\rightarrow 1$. \end{assumption}

The above condition is a high-level assumptions allowing us to handle an empirical process in the expansion of our statistic. It essentially introduces a Donsker feature for the function $\widehat{\varphi}$. Such an assumption has been used among others by \citet{rothe_semiparametric_2009}, \citet{escanciano_identification_2016}, and \citet{mammen_semiparametric_2016}. Sufficient conditions for it can be found in \citet{vaart_asymptotic_1998} or \citet{vaart_weak_1996}. We have also derived alternative proofs based on sample splitting or cross-fitting that avoid such a high-level condition. We have run simulations with cross-fitting and sample-splitting, but we did not notice any improvement in terms of finite sample performance. Thus, we choose to keep the above assumption and avoid sample splitting (or cross-fitting) for a simple implementation of the test. Alternatively  such a high-level condition could be avoided by using a Sobolev penalized method for estimating $\varphi$. This however would produce much longer proofs and complicate the implementation of the test.

Let $\P_n$ be the empirical mean operator. We have the following theorem.
\begin{theorem}\label{expansion of the statistic with NPIV}
Let Assumptions \ref{W} and \ref{completeness} to \ref{Donsker} hold. Under $H_0$,
\begin{align*}
S_n=& \sqrt{n}\mathbb{P}_n\,X\,(W-\mathbb{E}[W])\,U -\mathbb{E}[UX]\, \sqrt{n}\,\mathbb{P}_n\,(W-\mathbb{E}[W])\\
  &-\sqrt{n}\,\mathbb{P}_n\, U(A(A^*A)^{-1}g)(W)+o_P(1)\, .
\end{align*}
If instead $\mathbb{E}[\widetilde{U}X(W-\mathbb{E}[W])]\neq 0$, then 
$|S_n|/\sqrt{n}\to C\ne 0$.
\end{theorem}

The proof of the Theorem is in Section S2.1 of the supplementary material. Let us briefly comment on the asymptotic expansion (influence-function representation) 
of the statistic. The first term on the right hand side is the version of our statistic that we could use  had we observed the error $U$ and $\mathbb{E}[W]$. 
The second term arises because the expectation $\mathbb{E}[W]$ is unobserved
and is replaced by its estimated counterpart. Similarly, the third term arises because the error $U$ is unobserved
and to estimate it we replace the true function $\varphi$ with its estimator $\widehat{\varphi}$. More in detail, 
such a term originates from a (nonparametric) bias involving the difference $\widehat{\varphi}-\varphi$ that 
appears in the expansion of $S_n$. 
This inflates the asymptotic variance of the statistic, as compared to the case where $U$ is known, 
and hence represents the price to pay for not observing the error term. To handle this bias term, we rely on decompositions from \cite{darolles2011nonparametric} and \cite{babii2017unobservables}.\\
The expansion in Theorem \ref{expansion of the statistic with NPIV} allows us to obtain the asymptotic normality of our statistic under the null hypothesis. 
To this end, notice that the first term has zero expectation under the null of {homogeneous treatment effects}. The second term has 
clearly a null expectation. Finally, since $\mathbb{E}[U|W]=0$, the last term also has zero expectation. 
Therefore, a standard Central Limit Theorem implies that $S_n \overset{d}{\to} \mathcal{N}(0,\Psi)$, where $\Psi$  is a covariance matrix defined by the influence function representation in Theorem \ref{expansion of the statistic with NPIV}. Since such a covariance matrix has an intricate expression, it would be unconvenient to estimate it in practice. Thus, we suggest to bootstrap the statistic $S_n$ by the pairwise scheme to obtain the p-values of the test. As for the linear test, the fact that $|S_n|/\sqrt{n}\to C\ne 0$ when $\mathbb{E}[\widetilde{U}X(W-\mathbb{E}[W])]\neq 0$ holds implies that the power of the test goes to $1$ as $n$ goes to $\infty$ under alternative hypotheses satisfying $\mathbb{E}[\widetilde{U}X(W-\mathbb{E}[W])]\neq 0$. 

Notice that we are keeping Assumption \ref{completeness} both under the null of {homogeneous treatment effects} and under the alternative. We have chosen to do this to simplify the exposition. However, it is possible that if $\mathbb{E}[\widetilde U X (W-\mathbb{E}[W])]\neq 0$ (and {treatment effects are not homogeneous}), the NPIV model might be misspecified and/or completeness might not hold. In such a case we would need to modify the proof about the power of the test. Let us briefly discuss these modifications. First, when the model is misspecified and completeness does not hold, it is possible to show that the Tikhonov regularized estimator $\widehat{\varphi}$ converges to $\widetilde{\varphi}^{\perp}$ that is the element of $\mathcal{N}(A)^{\perp}$ (the orthogonal complement of the null space of $A$) that solves $\min_{\phi\in \mathcal{N}(A)^{\perp}}\|r-A\phi\|$.  In this case, we could define $\widetilde{U}$ as $Y-\widetilde{\varphi}^\perp (Z)$  and modify the statement of Theorem \ref{expansion of the statistic with NPIV} accordingly.

\subsection{Simulations}\label{subsec:simulations}
We set $\varphi(z):=\left(1+\exp(-z)\right)^{-2};\ X\sim\mathcal{N}(-0.5,1);\  W\sim\mathcal{N}(0,1);\ V\sim\mathcal{N}(0,1);\ Z=0.4 W + 0.2 V;  \ U(z)=(V+X)\,(1+\gamma z)\text{ for all } z\in\R,$
where $X,W,V$ are mutually independent. The outcome $Y$ is equal to $\varphi(Z)+U(Z)$. When $\gamma=0$ we are under the null of homogeneous treatment effects, while for $\gamma\neq 0$ we are under the alternative hypothesis. So, $\gamma$ represents the magnitude of the departure from the null. 

To check the robustness of our test with respect to the choices of the smoothing and regularization parameters, we let them vary around benchmark values. So, we set $(h_Z,h_W)=C_h (h^*_Z,h_W^*)$ and $\lambda=C_{\lambda} \lambda^*$, where $(h_Z^*,h_W^*)=(\text{sd}(Z)n^{-1/5}, \text{sd}(W)n^{-1/5})$, where $\text{sd}(Z)$ (resp. $\text{sd}(W)$) is the empirical standard deviation of $Z$ (resp. $W$), while $\lambda^*$ is the regularization parameter selected according to the Cross-Validation method, as seen in Section \ref{sec: implementation of NPIV statistic}..  Note that $(h_Z^*,h_W^*)$ are the bandwidths set according to a version of Silverman's Rule of Thumb (\cite{silverman1986density}). $C_h$ and $C_{\lambda}$ are fixed constants. We run simulations for $C_h,C_{\lambda}\in\{0.5,1,2\}$ and we consider the modest sample sizes of $n=100,250,500$. As a kernel we use the standard Gaussian density. We set $\pi$ to the Gaussian density with mean equal to the sample mean of $Z$ and variance equal to twice the sample variance of $Z$. Similarly, the measure $\tau$ is set to the Gaussian density with mean equal to the sample mean of $W$ and variance equal to twice the sample variance of $W$. \citet{florens_instrumental_2012} used a similar strategy to select $\pi$ and $\tau$ in their simulation setting. Intuitively, since $\pi$ and $\tau$ have to be two densities and at the same time appear at the denominators in 
$\widehat{A}$ and $\widehat{A}^*$, we want them (a) to integrate to 1 and 
(b) to converge towards zero sufficiently slowly on the tails. To speed up
computations, we use the warp-speed method by \citet{giacomini_warp-speed_2013}: for each Monte Carlo iteration we draw a single bootstrap
sample, and we use the whole set of bootstrap statistics to compute
the bootstrap p-values associated with each original statistic. We perform a very large number of Monte Carlos iterations equal to 10,000. 

The results under the null hypothesis of {homogeneous treatment effects} ($\gamma=0$) are reported in Table \ref{table for size}. The tests are implemented at 5 and 10 percent nominal levels. The nominal sizes of each of the tests are in bold. The result show that the test is reasonably stable with respect to the choices of $\lambda$, $h_Z$, and $h_W$ and controls well the size under the null hypothesis. Also, as expected, the error in the rejection probability (i.e. the difference between the empirical rejection proportions and the nominal size of the test) tends to become smaller as the sample size increases.

\begin{figure}[ht]
\begin{center}
\includegraphics[width=9cm,height=9cm]{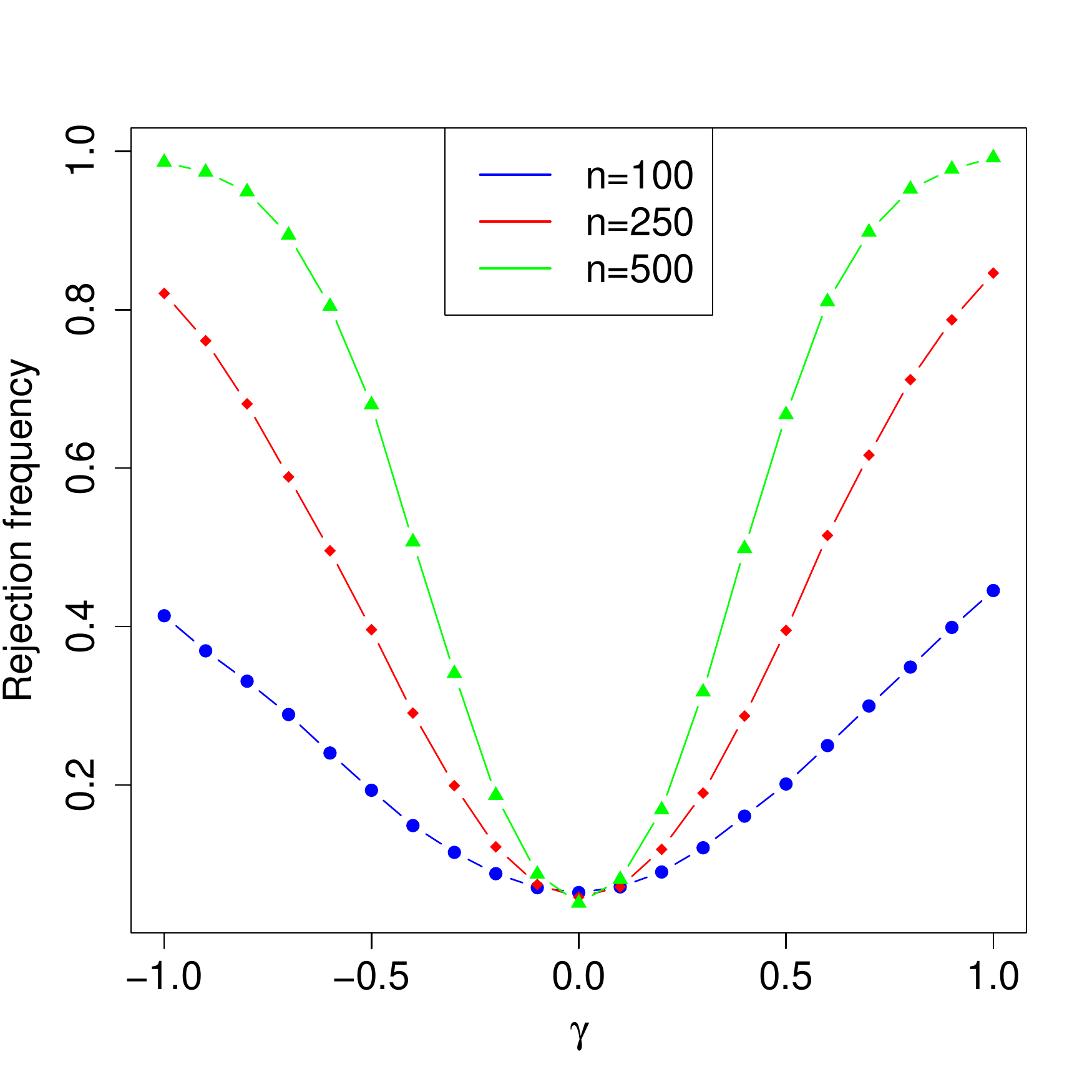}
\end{center}
\caption{Power curve for 5\% level tests.}
\label{Fig}
\end{figure}

To check the power properties, we run simulations under the alternative hypothesis for different values of  $\gamma$. The results are shown in Figure \ref{Fig}. We only report results for the benchmark values of $h_Z$, $h_W$, and $\lambda$ and for the nominal size of 5 percent. For the other choices of $h_Z$, $h_W$, and $\lambda$ and for the 10 percent nominal level the results are qualitatively similar. The test shows good power under the alternative hypothesis. As the departure from the null increases, i.e. $\gamma$ gets further from 0, the test rejects the null hypothesis with an increasing frequency. Also, such a rejection frequency increases with the sample size at every value of $\gamma$. 

To sum up, these simulations show that (a) for the moderate sample sizes of $n=100,250,500$ the tests display a good performance both in terms of size control and in terms of power, and (b) the results are stable with respect to the choices of the smoothing and regularization parameters. 

\begin{table}[ht]
\centering
\renewcommand{\arraystretch}{0.8} %
\begin{tabular}{rrrrrrrr}

\hline
& &\multicolumn{2}{c}{$\lambda^*$} & \multicolumn{2}{c}{0.5 $\lambda^*$} & \multicolumn{2}{c}{2 $\lambda^*$} \\

  \hline
 & &\textbf{0.05} & \textbf{0.10} & \textbf{0.05} & \textbf{0.10} & \textbf{0.05} & \textbf{0.10} \\ 
  \hline
  $n$=100 \hspace{2cm}&  $h^*$ & 0.0642 & 0.1250 & 0.0635 & 0.1246& 0.0656& 0.1227 \\ 
&  0.5 $h^*$ & 0.0803 & 0.1465 & 0.0800 & 0.1461& 0.0761 & 0.1500 \\ 
&2 $h^*$ & 0.0618& 0.1240 & 0.0612 & 0.1201 & 0.0619& 0.1277 \\
  \hline
 $n$=250\hspace{2cm}&$h^*$ & 0.0583 & 0.1136 & 0.0553 & 0.1101 & 0.0592& 0.1139\\ 
& 0.5 $h^*$ & 0.0640& 0.1246 & 0.0661& 0.1228 & 0.0634& 0.1247 \\ 
& 2 $h^*$ & 0.0557 & 0.1155 & 0.0553 & 0.1111 & 0.0577 & 0.1163 \\ 
  \hline
  $n$=500 \hspace{2cm}&$h^*$ & 0.0509 & 0.1070 & 0.0511& 0.1046 & 0.0546 & 0.1091\\ 
& 0.5 $h^*$ & 0.0568 & 0.1057 & 0.0556 & 0.1074 & 0.0568 & 0.1072 \\ 
 &2 $h^*$ & 0.0457 & 0.0991 & 0.0449 & 0.0964& 0.0496& 0.1011 \\ 
   \hline
\end{tabular}
\caption{Empirical rejections of the tests under the null hypothesis $\gamma=0$.  }
\label{table for size}
\end{table}

\subsection{Application to real data}\label{sec.emp}

In this section we apply the test of {homogeneous treatment effects} to a real dataset. Note that the application of returns to schooling is not well-suited to the nonparametric test because, in this application, the endogenous variable (education) has a support much larger than the instrument (college proximity) which is binary. 

Instead, we focus on estimating a fish demand equation.This describes the relationship between the demanded quantity of fish and its price. It is common practice in econometrics to assume that prices are endogenous and estimate demand equations by instrumental variable regressions. 

The data we use come from \citet{graddy2006markets} and contain daily observations at the New York Fulton fish market about the log of the price, the log of the sold quantity, an indicator for each day of the week, and the wind speed registered in the previous day. The dataset can be downloaded at “\url{https://users.ox.ac.uk/~nuff0078/EconometricModeling/}”. The total number of observations is 111. 

We study a demand equation. The outcome $Y$ is equal to the logarithm of the quantity sold and $Z$ is the logarithm of the market price. Market price is likely to be endogenous since it also depends on expected demand, which itself also affects the quantity. Following the literature (\citet{ graddy2006markets}), we choose a variable related to weather as an instrument, so we set $W$ equal to the wind speed recorded on the day corresponding to the observation. Such a variable is viewed as sufficiently correlated with the price ($Z$), since the weather affects the ability to fish and is at the same time exogenous with respect to the errors $\{U(z)\}_z$ because it is probably unrelated to demand shocks related to human factors.

The choice of the covariable $X$ is crucial. We set $X$ equal to the indicator for week days (Monday to Thursday). Such a variable should not be correlated with the wind speed since the weather does not depend on the day of the week. In fact, a Kolmogorov-Smirnov test does not reject the hypothesis of independence between $X$ and $W$ (p-value of $0.17$).  Hence, we can reasonably assume that $(\{U(z)\}_z,X)\independent W_k,$ which corresponds to Assumption \ref{W}.

We perform the nonparametric test using the cross-validated penalty $\lambda^*$ and bandwidth $(h_Z^*,h_W^*)=(\text{sd}(Z)n^{-1/5}, \text{sd}(W)n^{-1/5})$, where $\text{sd}(Z)$ (resp. $\text{sd}(W)$) is the empirical standard deviation of $Z$ (resp. $W$) and $n$ is the sample size.  This is a version of Silverman's Rule of Thumb (\cite{silverman1986density}). We use $1,000$ bootstrap resamplings. The p-value of the test is $0.0362$. Hence, we reject the null hypothesis of {homogeneous treatment effects} at the 5 percent nominal level. So, we conclude that the causal intepretation of the separable NPIV regression should be doubted for this specific example. This is a reason to use a different approach such as instrumental variable quantile regression (\cite{chernozhukov2008instrumental}) which allows heterogeneity of treatment effects by quantiles.

\section*{Supporting information}
All the proofs along with some examples regarding the analysis of the power of the nonparametric test can be found in the online supplement.

\bibliographystyle{chicago}
\bibliography{paper}


\end{document}



\title{\bf Online supplement to ``Testing for homogeneous treatment effects in linear and nonparametric instrumental variable models''}
\author{
{\large Jad B\textsc{eyhum}}
\footnote{ORSTAT, KU Leuven. Financial support from the European Research Council (2016-2021, Horizon 2020 / ERC grant agreement No.\ 694409) is gratefully acknowledged.}\\\texttt{\small jad.beyhum@kuleuven.be}
\and
\addtocounter{footnote}{2}
{\large Jean-Pierre F\textsc{lorens}}
\footnote{Toulouse School of Economics, Universit\'e Toulouse Capitole. Jean-Pierre Florens acknowledges funding from the French National Research Agency (ANR) under the Investments for the Future program (Investissements d'Avenir, grant ANR-17-EURE-0010).}
\\\texttt{\small jean-pierre.florens@tse-fr.eu}
\and {\large Elia L\textsc{apenta}}
\footnote{CREST, ENSAE Paris, Institut Polytechnique de Paris}
\\\texttt{\small elia.lapenta@gmail.com}
\and
{\large Ingrid V\textsc{{an} K{eilegom}}\ $^*$}
\\\texttt{\small ingrid.vankeilegom@kuleuven.be}
}

\maketitle

\spacingset{1.4}


Section \ref{lin} is concerned with the linear test. It contains the proofs of a result regarding Example 1 (\ref{lin.ex}) and of Theorem 2.1 (\ref{lin.th}). Section \ref{npt} treats the nonparametric test. It includes the proof of the results of Section 3 in the main text (\ref{npt.th}) and a theoretical analysis of the power of the test through examples (\ref{app.C}).

 \section{On the linear test}\label{lin}\setcounter{equation}{0}

\subsection{On example 1}\label{lin.ex}
Let us show that $\beta_2^{OLS}=\alpha/3$ in this example. We write \begin{equation}\label{ex1linear}Y=\E[Y|Z_2=0]+Z_2(\E[Y|Z_2=1]-\E[Y| Z_2=0])  + V.\end{equation} It holds that 
\begin{align*}
\E[V]&= \E[Y-\E[Y|Z_2=0]-Z_2(\E[Y|Z_2=1]-\E[Y|Z_2=0])]\\
&= \E[Y-\E[Y|Z_2=0]|Z_2=0]\P(Z_2=0)+ \E[Y-\E[Y| Z_2=1]| Z_2=1]\P( Z_2=1)=0
\end{align*}
and 
\begin{align*}
\E[Z_2 V]&= \E[Z_2\{Y-\E[Y|Z_2=0]-Z_2(\E[Y|Z_2=1]-\E[Y|Z_2=0])\}]\\
&= \E[Y-\E[Y|Z_2=1]|Z_2=1]\P(Z_2=1)=0.
\end{align*}
Hence, \eqref{ex1linear} is a linear projection of $Y$ on $Z$. Since this linear projection is unique (because $E[Z_2^2]>0$), we get 
\begin{align*}
\beta_1^{OLS}&= \E[Y|Z_2=0]\\
\beta_2^{OLS}&=\E[Y|Z_2=1]-\E[Y|Z_2=0].
\end{align*}
Now, let us compute $\beta_2^{OLS}$. We have $\E[Y|Z_2=0]=0$ because $U(Z)=Z_2\left(1\left\{\epsilon\ge \frac34\right\}-\frac14\right)\alpha$ and 
\begin{align*}\E[Y|Z_2=1]&=\alpha \P\left(\left.\epsilon\ge \frac34\right|Z_2=1\right)\\
&= \alpha \frac{\P\left(\left\{\epsilon\ge \frac34\right\}\cap \{Z_2=1\}\right)}{\P(Z_2=1)}\\
&=\alpha \frac{\P\left(\left\{\epsilon\ge \frac34\right\}\cap \{W_2=1\}\right)}{3/8}=\alpha\frac{1/8}{3/8}=\frac{\alpha}{3}.\end{align*}

\subsection{Proof of Theorem 2.1}\label{lin.th}
Theorem 1 is a direct corollary of the following lemma, which gives the influence function representation of the test statistic.
\begin{lemma} \label{ae}Let Assumptions 2.1 and 2.2 in the main text hold. Then, we have$$T_n=\Psi \frac{1}{\sqrt{n}}\left(\sum_{i=1}^nV_i\right)+ o_P\left(\frac{1}{\sqrt{n}}\right),$$
where
\begin{align*}V_i=\left(\begin{array}{c} U_i^{TSLS}X_i\widetilde{W}_{ki} \\ \widetilde{W}_{ki} \\  U_i^{TSLS}W_{i} \end{array}\right);\qquad \Psi= \left(I_{q}, -\E[U^{TSLS}X]I_{q},-\E[X\widetilde{W}_kZ^\top]\Sigma^{TSLS} \right);\\
\Sigma^{TSLS}=\left(\E[ZW^\top] \E[WW^\top]^{-1}\E[ZW^\top]^\top\right)^{-1}\E[ZW^\top]\E[WW^\top]^{-1}; \qquad \widetilde{W}=W-E[W].
\end{align*}
\end{lemma}
\begin{Proof}
Since $\widehat{U}_i=Y_i-Z_i^\top\widehat \beta^{TSLS}=U_i^{TSLS}-Z_i^\top(\widehat \beta^{TSLS}-\beta^{TSLS})$, it holds that
\begin{align}\notag &\frac1n \sum_{i=1}^n  \widehat U_iX_i(W_{ki}-\overline{W}_k)\\
\label{2tustat} &= \frac1n \sum_{i=1}^n  U_i^{TSLS}X_i(W_{ki}-\overline{W}_k)
-\left(\frac{1}{n} \sum_{i=1}^n   X_i(W_{ki}-\overline{W}_k)Z_i^\top\right)(\widehat \beta^{TSLS}-\beta^{TSLS}).
\end{align}
We deal with the first term on the right-hand side of \eqref{2tustat}. We have
\begin{align*}&\frac1n \sum_{i=1}^n  U_i^{TSLS}X_i(W_{ki}-\overline{W}_k)\\
&=\frac1n \sum_{i=1}^n  U_i^{TSLS}X_i\widetilde W_{ki}+\frac1n \sum_{i=1}^n  U_i^{TSLS}X_i(E[W_k]-\overline{W}_k)\\
&=\frac1n \sum_{i=1}^n  U_i^{TSLS}X_i\widetilde W_{ki}-\left[\frac1n \sum_{i=1}^n  U_i^{TSLS}X_i\right] (\overline{W}_k -E[W_k])\\
&=\frac1n \sum_{i=1}^n  U_i^{TSLS}X_i\widetilde W_{ki}- \E[U^{TSLS}X]\left[\frac1n\sum_{j=1}^n \widetilde W_{ki}\right]+o_P\left(\frac{1}{\sqrt{n}}\right).
\end{align*}
Then, we handle the second term on the right-hand side of \eqref{2tustat}. It holds that
\begin{align*}&\frac1n \sum_{i=1}^n  X_i(W_{ki}-\overline{W}_k)Z_i^\top\\
&=\frac1n \sum_{i=1}^n  X_i\widetilde W_{ki}Z_i^\top+\frac1n \sum_{i=1}^n  X_i(E[W_k]-\overline{W}_k)Z_i^\top\\
&=\frac1n \sum_{i=1}^n  X_i\widetilde W_{ki}Z_i^\top -(E[W_k]-\overline{W}_k)\left[\frac1n \sum_{i=1}^n  X_iZ_i^\top\right].
\end{align*}
Since $\widehat\beta^{TSLS}-\beta=O_P(1/\sqrt{n})$, and by the law of large numbers 
$$(E[W_k]-\overline{W}_k)\left[\frac1n \sum_{i=1}^n  X_iZ_i^\top\right]=o_P(1), $$
we get 
\begin{align*}\left(\frac{1}{n} \sum_{i=1}^n   X_i(W_{ki}-\overline{W}_k)Z_i^\top\right)(\widehat \beta^{TSLS}-\beta)
&=\frac1n \sum_{i=1}^n  X_i\widetilde W_{ki}Z_i^\top(\widehat \beta^{TSLS}-\beta) +o_P\left(\frac{1}{\sqrt{n}}\right)\\
&=\E[X(W_k-\E[W_k])Z^\top](\widehat \beta^{TSLS}-\beta)+o_P\left(\frac{1}{\sqrt{n}}\right).
\end{align*}
Moreover, by standard computations, we have 
\begin{align*}\widehat{\beta}^{TSLS}
&=\Sigma^{TSLS} \left(\frac1n\sum_{i=1}^nU_i^{TSLS}W_i\right)+\beta + o_P\left(\frac{1}{\sqrt{n}}\right),
\end{align*}
which leads to the result.

\end{Proof}

\section{On the nonparametric test}\label{npt}

\subsection{Proof of the results of Section 3}\label{npt.th}
Recall that $\mathbb{P}_n$ represents the empirical mean operator. Let us denote with $\P$ the population mean operator, so $\P f(\xi)=\E f(\xi)$ for any random variable $\xi$. We finally denote with $\|\cdot\|_{\pi}$ the norm of the $L^2_{\pi}(\R)$ space, i.e. $\|f\|^2_\pi=\int|f(z)|^2\pi(d\,z)$ for any $f\in L^2_{\pi}(\R)$.

\subsubsection{Proof of Theorem 3.1}
Let us decompose the statistic as follows:
\begin{align}\label{decomposition of the NPIV statistic}
    \sqrt{n}\mathbb{P}_n \widehat{U}X(W-\overline{W})=& \sqrt{n}\mathbb{P}_n U X(W-\mathbb{E}[W])\nonumber \\
    &+ \sqrt{n}(\mathbb{P}_n-\P)(\varphi-\widehat{\varphi})X(W-\mathbb{E}[W]) \nonumber \\
    &+ \sqrt{n}\P(\varphi-\widehat{\varphi})X(W-\mathbb{E}[W]) \nonumber \\
    &- \sqrt{n}\mathbb{P}_n\widehat{U}X(\overline{W}-\mathbb{E}[W])\, . 
\end{align}

We start by showing that the last term on the RHS equals $\mathbb{E}[U X]\cdot\sqrt{n}\mathbb{P}_n(W-\mathbb{E}[W])+o_P(1)$. Notice that  $\sqrt{n}\mathbb{P}_n\widehat{U}X(\overline{W}-\mathbb{E}[W])=\left[\mathbb{P}_n\widehat{U}X\right]\cdot\sqrt{n}(\mathbb{P}_n-\P)W$. By Assumption 3.6 in the main text, with probability approaching one
\begin{align*}
    \left|(\mathbb{P}_n-P)\widehat{U}X\right|\leq \sup_{\phi\in \Upsilon}\left|(\mathbb{P}_n-\P)\phi \right|\, ,
\end{align*}
where  $\Upsilon:=\{(u,z,x)\mapsto (y-\phi(z))\,x\,:\,\phi\in\Phi\}$. It follows from the definition of bracketing number that $N_{[\,\,]}(\epsilon,\Upsilon,\|\cdot \|_{\P})\leq N_{[\,\,]}(\mathbb{E}[X^2]\cdot\epsilon,\Phi,\|\cdot \|_{\P}) $. Assumption 3.6 ensures that the RHS of this inequality is finite for any $\epsilon$. So, the class $\Upsilon$ is Glivenko-Cantelli and by Theorem 19.4 in \citet{vaart_asymptotic_1998} we get that $(\mathbb{P}_n-\P)\phi=o_P(1)$ uniformly in $\phi\in \Upsilon$. This and the previous display lead to 
 \begin{align*}
     \mathbb{P}_n \widehat{U}X=\P\,U\,X\,+\P\,(\varphi-\widehat{\varphi})\,X\,+\,o_P(1)\, .
 \end{align*}
To show the negligibility of the second term on the RHS, by the Law of Iterated Expectations and the Cauchy-Schwartz inequality we get 
\begin{align*}
    \left|\P(\widehat{\varphi}-\varphi)X\right|^2=& \left| \int_{ }\,(\widehat{\varphi}(z)-\varphi(z))\, \mathbb{E}[X|Z=z]\,\frac{f_{Z}(z)}{\pi(z)}\,\pi(dz)\,  \right| \\
    \leq & \|\widehat{\varphi}-\varphi\|^2_{\pi}\cdot \|\mathbb{E}[X|Z=\cdot]f_Z/\pi\|_{\pi}^2 \, 
\end{align*}
where $\|\cdot \|_{\pi}$ denotes the norm on $L^2_{\pi}(\mathbb{R})$. The second factor on the RHS is finite by Assumption 3.2 in the main text. The first factor is instead $o_P(1)$ by Lemma \ref{Lem: rate for phihat}. By putting together these results, we find that
\begin{equation}\label{expansion 4th term for the statistic}
\sqrt{n}\mathbb{P}_n\widehat{U}X(\overline{W}-\mathbb{E}[W])=\mathbb{E}[U X]\cdot \sqrt{n}\mathbb{P}_n(W-\mathbb{E}[W])+o_P(1)\, .    
\end{equation}

We now show the negligibility of the second term on the RHS of Equation (\ref{decomposition of the NPIV statistic}). Combining the Law of Iterated Expectations and Assumption 3.2 leads to 
\begin{align*}
    \|(\widehat{\varphi}-\varphi)X(W-\mathbb{E}[W])\|_P^2=&\int_{ }\left|\widehat{\varphi}(z)-\varphi(z)\right|^2\, \frac{\mathbb{E}[X^2(W-\mathbb{E}[W])^2|Z=z]f_Z(z)}{\pi(z)}\, \pi(dz) \\
    \leq & C\|\widehat{\phi}-\phi_0\|^2_\pi \, ,
\end{align*}
with $\|\widehat{\varphi}-\varphi\|^2_\pi=o_P(1)$, as previously found. So, by Assumption 3.6 and the boundedness of  $\mathbb{E}[X^2(W-\mathbb{E}[W])^2|Z=\cdot]$ (see Assumption 3.2), the conditions of Lemma \ref{ASE lemma} are satisfied and we obtain
\begin{equation}\label{expansion 2nd term for the statistic}
    \sqrt{n}(\mathbb{P}_n-\P)(\widehat{\varphi}-\varphi)\,X\,(W-\mathbb{E}[W])=o_P(1)\, .
\end{equation}
Gathering together Equations (\ref{decomposition of the NPIV statistic}), (\ref{expansion 2nd term for the statistic}), (\ref{expansion 4th term for the statistic}),  and the result of Lemma \ref{exansion of the inner product} ahead delivers the desired result under the null hypothesis of error invariance.\\
To show the behavior of the statistic under the alternative, we can proceed as at the beginning of this proof to obtain 
\begin{equation*}
    \mathbb{P}_n \widehat{U}X(W_k-\mathbb{E}[W_k])=\mathbb{E}[\widetilde{U}X(W-\mathbb{E}[W])]+o_P(1)
\end{equation*}
where $\widetilde{U}=Y-\widetilde{\varphi}(Z)$ and the first leading term is different from $0$ under $H_1$.

\subsubsection{Auxiliary lemma}
In this section we provide an auxiliary lemma we used in the proofs of Theorem 3.1. Let us denote with $\left<\cdot,\cdot\right>$ the inner product on either $L^2_{\pi}(\R)$ or $L^2_{\tau}(\R)$, and let $\|\cdot\|$ denote the norm induced by such an inner product. Whether $\left<\cdot,\cdot\right>$ and  $\|\cdot\|$ refer to either $L^2_{\pi}(\R)$ or $L^2_{\tau}(\R)$ will be clear by their arguments. Also, given the operators $A$ and $A^*$ we define
\begin{align*}
    \|A\|^2_{op}:=\sup_{\phi\in L^2_{\pi}(\R):\|\phi\|=1}\|A\phi\|^2\quad \text{ and }\quad \|A^*\|^2_{op}:=\sup_{\psi\in L^2_{\tau}(\R):\|\psi\|=1}\|A^*\psi\|^2\, .
\end{align*}
Since $A$ and $A^*$ are linear bounded operators, both  $\|A\|^2_{op}$ and $\|A^*\|^2_{op}$ are finite. Also, since $A^*$ is the Hilbert adjoint of $A$, it holds that $\|A\|^2_{op}$ and $\|A^*\|^2_{op}$ are equal. 

\begin{lemma}\label{exansion of the inner product}
Under Assumptions 3.1 to 3.6 in the main text, we have
$$\sqrt{n}\, \P\, (\widehat{\varphi}-\widetilde{\varphi})\,g=\sqrt{n}\mathbb{P}_n\, \widetilde{U} \, \left[ A(A^*A)^{-1}g \right](W)+o_P(1)\, . $$
\end{lemma}

\begin{Proof} Let us start with the following decomposition

\begin{equation}\label{eq: decomposition of phi.hat}
\widehat{\varphi}-\widetilde{\varphi}= \Xi_1 + \Xi_2 + \Xi_3 + \Xi_4 + (\widetilde{\varphi}_\lambda - \widetilde{\varphi}) 
\end{equation}

where 

\begin{align*}
    \Xi_1:=& (\lambda I + A^*A)^{-1}A^* (\widehat{r}-\widehat{A}\widetilde{\varphi}) \\
    \Xi_2 := & (\lambda I + A^*A)^{-1}(\widehat{A}^*-A^*)(\widehat{r}-\widehat{A}\widetilde{\varphi})\\
    \Xi_3 := & \left[ (\lambda I + \widehat{A}^*\widehat{A})^{-1} -(\lambda I + A^*A)^{-1}  \right]\widehat{A}^*(\widehat{r}-\widehat{A}\widetilde{\varphi})\\
    \Xi_4:= & (\lambda I + \widehat{A}^*\widehat{A})^{-1}\widehat{A}^*\widehat{A}(\widetilde{\varphi} - \widetilde{\varphi}_\lambda)\\
    \widetilde{\varphi}_\lambda := & (\lambda I + A^*A)^{-1}A^*r\text{ . }
\end{align*}
The previous decomposition holds because  
\begin{align*}
\Xi_1 + \Xi_2 + \Xi_3 =  (&\lambda I + A^*A)^{-1}A^*(\widehat{r}-\widehat{A}\widetilde{\varphi})\\
+ &  (\lambda I + A^*A)^{-1}\widehat{A}^*(\widehat{r}-\widehat{A}\widetilde{\varphi})- (\lambda I + A^*A)^{-1}A^*(\widehat{r}-\widehat{A}\widetilde{\varphi}) \\
+ & (\lambda I + \widehat{A}^*\widehat{A})^{-1}\widehat{A}^*(\widehat{r}-\widehat{A}\widetilde{\varphi}) - (\lambda I + A^*A)^{-1}\widehat{A}^*(\widehat{r}-\widehat{A}\widetilde{\varphi})\\
= & \widehat{\varphi}- (\lambda I + \widehat{A}^*\widehat{A})^{-1}\widehat{A}^*\widehat{A}\widetilde{\varphi} \, .
\end{align*}
Now, 
\begin{align*}
    \left| \left< \Xi_2 , g \right> \right|=& \left|\left< (\widehat{A}^*-A^*)(\widehat{r}-\widehat{A}\widetilde{\varphi}) ,(\lambda I + A^*A)^{-1} g \right>\right|\\
    \leq & \left| \left| \widehat{A}-A \right| \right|_{op} \cdot \left| \left| \widehat{r}-\widehat{A}\widetilde{\varphi} \right| \right|\cdot \left| \left| (\lambda I + A^*A)^{-1}g \right| \right| \\
    \leq & C   \left| \left| \widehat{A}-A \right| \right|_{op} \cdot \left| \left| \widehat{r}-\widehat{A}\widetilde{\varphi} \right| \right| \lambda^{\frac{\eta \wedge 2 }{2} - 1} \text{ , }
\end{align*}
where the second line follows from the Cauchy-Schwartz inequality and the continuity of $\widehat{A}-A$, while the third line follows from Assumption 3.5(a) and Lemma \ref{Lemm: Bounds on Compact Operator}(d). Assumption 3.5(a) ensures that $\eta\geq 2$. As already noticed before, $||\widehat{A}-A||_{op}/\sqrt{\lambda}=o_P(1)$ and $||\widehat{r}-\widehat{A}\widetilde{\varphi}||/\sqrt{\lambda}=o_P(1)$. So, by the above display and Assumption 3.4 we find 
\begin{equation}\label{eq: Xi2}
    \sqrt{n}\left<  \Xi_2 , g \right>=O_P(\sqrt{n\lambda^2})=o_P(1)\text{ . }
\end{equation}
We now handle the term $\sqrt{n}\left< \widetilde{\varphi}_\lambda - \widetilde{\varphi},g \right>$. By Assumption 3.5(b) and Lemma \ref{Lemm: Bounds on Compact Operator}(f) $||\widetilde{\varphi}_\lambda - \widetilde{\varphi}||=O(\lambda^{\frac{\theta \wedge 2}{2}})$ with $\theta\geq 2$.  Combining this rate with the Cauchy-Schwartz inequality gives 
\begin{equation}\label{eq: inner product of Regularization Bias }
    \sqrt{n}\left|\left< \widetilde{\varphi}_\lambda - \widetilde{\varphi} , g \right> \right|\leq \sqrt{n}\cdot||\widetilde{\varphi}_\lambda - \widetilde{\varphi} ||\cdot ||g||=O(\sqrt{n\lambda^2})=o(1)\text{ . }
\end{equation}
To show the negligibility of $\sqrt{n}\left< \Xi_3 , g \right>$, notice that 
\begin{align*}
    \Xi_3= & (\lambda I + A^*A)^{-1} \left[  (\lambda I + A^*A) - (\lambda I + \widehat{A}^*\widehat{A}) \right](\lambda I + \widehat{A}^*\widehat{A})^{-1}\widehat{A}^*(\widehat{r}-\widehat{A}\widetilde{\varphi})\\
    = & (\lambda I + A^*A)^{-1}\left[ A^*(A-\widehat{A})+(A^*-\widehat{A}^*)\widehat{A} \right](\lambda I + \widehat{A}^*\widehat{A})^{-1}\widehat{A}^*(\widehat{r}-\widehat{A}\widetilde{\varphi})\\
    = & (\lambda I + A^*A)^{-1}A^*(A-\widehat{A})(\lambda I + \widehat{A}^*\widehat{A})^{-1}\widehat{A}^*(\widehat{r}-\widehat{A}\widetilde{\varphi})\\
    \text{ }&+ (\lambda I + A^*A)^{-1}(A^*-\widehat{A}^*)\widehat{A}(\lambda I + \widehat{A}^*\widehat{A})^{-1}\widehat{A}^*(\widehat{r}-\widehat{A}\widetilde{\varphi})\text{ . }
\end{align*}
In view of the above decomposition and the Cauchy-Schwartz inequality
\begin{align*}
    \sqrt{n}\left|\left< \Xi_3 , g \right>\right|\leq & \sqrt{n}\left|\left< \widehat{r}-\widehat{A}\widetilde{\varphi} \text{ },\text{ } \widehat{A}(\lambda I + \widehat{A}^*\widehat{A})^{-1}(A^*-\widehat{A}^*)A(\lambda I + A^*A)^{-1} g \right>\right|\\
    & + \sqrt{n}\left|\left< \widehat{r}-\widehat{A}\widetilde{\varphi} \text{ },\text{ }\widehat{A}(\lambda I + \widehat{A}^*\widehat{A})^{-1}\widehat{A}^*(A-\widehat{A})(\lambda I + A^*A)^{-1}g \right>\right|\\
    \leq &  \sqrt{n}\left|\left| \widehat{r}-\widehat{A}\widetilde{\varphi} \right|\right|\cdot \left|\left| \widehat{A}(\lambda I + \widehat{A}^*\widehat{A})^{-1}(A^*-\widehat{A}^*)A(\lambda I + A^*A)^{-1}g \right|\right|\\
    & + \sqrt{n}\left|\left|  \widehat{r}-\widehat{A}\widetilde{\varphi}  \right|\right|\cdot \left|\left| \widehat{A}(\lambda I + \widehat{A}^*\widehat{A})^{-1}\widehat{A}^*(A-\widehat{A})(\lambda I + A^*A)^{-1}g  \right|\right|\\
    \leq & \sqrt{n}\cdot ||\widehat{r}-\widehat{A}\widetilde{\varphi}||\cdot ||\widehat{A}(\lambda I + \widehat{A}^*\widehat{A})^{-1}||_{op}\cdot ||\widehat{A}-A||_{op}\cdot ||A(\lambda I + A^*A)^{-1}g|| \\
    & + \sqrt{n}\cdot ||\widehat{r}-\widehat{A}\widetilde{\varphi}||\cdot ||\widehat{A}(\lambda I + \widehat{A}^*\widehat{A})^{-1}\widehat{A}^*||_{op}\cdot ||\widehat{A}-A||_{op}\cdot ||(\lambda I + A^*A)^{-1}g||\text{ . }
\end{align*}
Lemma \ref{Lemm: Bounds on Compact Operator}(a) and (c) imply that $||\widehat{A}(\lambda I + \widehat{A}^*\widehat{A})^{-1}\widehat{A}^*||_{op}\leq 1$ and  $||\widehat{A}(\lambda I + \widehat{A}^*\widehat{A} )^{-1}||_{op}\lesssim 1/\sqrt{\lambda}$. Combining Assumption 3.5(a)  and Lemma \ref{Lemm: Bounds on Compact Operator}(d) and (e) gives  $||(\lambda I + A^*A)^{-1}g||\lesssim \lambda^{\frac{\eta \wedge 2}{2}-1}$ and $||A(\lambda I + A^*A)^{-1}g||\lesssim \lambda^{\frac{\eta+1}{2} \wedge 1 - 1} $ with $\eta\geq 2$. Finally, Assumption 3.4 and Lemma \ref{Lem: rates for Ahat and rhat} ensure that $||\widehat{A}-A||_{op}\lambda^{-3/4}=o_P(1)$ and  $||\widehat{r}-\widehat{A}\widetilde{\varphi}||\lambda^{-3/4}=o_P(1)$. Hence, by these rates and the previous display we obtain
\begin{equation}\label{eq: _Xi3}
\sqrt{n}\left|\left< \Xi_3 , g \right>\right|\leq \sqrt{n}\text{ }o_P(\lambda)=o_P(1)\text{ . }
\end{equation}

\noindent To show that $\sqrt{n}\left< \Xi_4 ,g \right>$ is negligible, we first define the operator 
\begin{equation}\label{eq: Astar A eta}
 f\mapsto (A^*A)^{\eta/2}f:=\sum_j\lambda_j^{\eta}\left< f,\varphi_j \right>\varphi_j\text{ , }(A^*A)^{\eta/2}:L^2(\mathbb{R})\mapsto L^2(\mathbb{R}) \text{ , }   
\end{equation}
and let 
\begin{equation}\label{eq: ftilde}
    \widetilde{f}:=\sum_j \lambda_j^{-\eta}\left< g , \varphi_j \right>\varphi_j\text{ . }
\end{equation}
Notice that by Assumption 3.5, $||\widetilde{f}||<\infty$ for some $\eta\geq 2$. Also, $(A^*A)^{\eta/2}\widetilde{f}=g$. By using these for $\eta=2$ we obtain 

\begin{align}\label{eq: inner product of xi4}
    \sqrt{n}\left| \left< \Xi_4, g \right> \right|  = &  \sqrt{n} \left| \left< \Xi_4, (A^*A)^{\eta /2} \widetilde{f} \right> \right| \nonumber \\
    = & \sqrt{n}\left| \left< \Xi_4 , (A^*A)\widetilde{f} \right> \right| \nonumber  \\
    \leq & \sqrt{n} \left| \left< \Xi_4, (A^*-\widehat{A}^*)A\widetilde{f}\right> \right| + \sqrt{n}\left| \left< \Xi_4, \widehat{A}^*A \widetilde{f} \right> \right| \nonumber \\
    = &  \sqrt{n} \left| \left< \Xi_4, (A^*-\widehat{A}^*)A\widetilde{f}\right> \right| + \sqrt{n}\left| \left< \widehat{A} \Xi_4, A \widetilde{f} \right> \right| \nonumber \\
    \leq & \sqrt{n}\cdot||\Xi_4||\cdot ||\widehat{A}-A||_{op}\cdot ||A\widetilde{f}||+ \sqrt{n}\cdot||\widehat{A}\Xi_4||\cdot||A\widetilde{f}||\text{ , }
\end{align}
where the second equality follows from $(A^*A)^{\eta/2}=A^*A$ for $\eta=2$.
Now, to bound the RHS we decompose $\Xi_4$ as 
\begin{align}\label{eq: decomposition for xi4}
    \Xi_4 = & (\lambda I + \widehat{A}^*\widehat{A})^{-1}\widehat{A}^*\widehat{A}\widetilde{\varphi}- (\lambda I + A^*A)^{-1}A^*A\widetilde{\varphi} \nonumber \\
    = & \left[ (\lambda I + \widehat{A}^*\widehat{A})^{-1}(\lambda I + \widehat{A}^*\widehat{A} - \lambda I)- (\lambda I + A^*A)^{-1}(\lambda I + A^*A - \lambda I )\right]\widetilde{\varphi}  \nonumber \\
    = & \lambda \left[  (\lambda I + A^*A)^{-1} - (\lambda I + \widehat{A}^*\widehat{A})^{-1} \right]\widetilde{\varphi}  \nonumber \\
    = & \lambda (\lambda I + \widehat{A}^*{A})^{-1}\left[ (\lambda I + \widehat{A}^*\widehat{A})    -(\lambda I + A^*A) \right](\lambda I + A^*A)^{-1}\widetilde{\varphi} \nonumber \\
    = &  \lambda (\lambda I + \widehat{A}^*{A})^{-1}\left[ \widehat{A}^*(\widehat{A}-A) +   (\widehat{A}^*-A^*)A  \right](\lambda I + A^*A)^{-1}\widetilde{\varphi} \nonumber \\
    = & \lambda(\lambda I + \widehat{A}^*\widehat{A})^{-1}\widehat{A}^*(\widehat{A}-A)(\lambda I + A^*A)^{-1}\widetilde{\varphi} \nonumber \\
    & + \lambda (\lambda I + \widehat{A}^*\widehat{A})^{-1}(\widehat{A}^*-A^*)A(\lambda I + A^*A)^{-1}\widetilde{\varphi}\text{ . }
\end{align}

By the above decomposition, we find that 
\begin{align}\label{eq: bound for Xi4}
    ||\Xi_4||\leq & ||\lambda (\lambda I + \widehat{A}^*A)^{-1} ||_{op}\cdot ||\widehat{A}^*||_{op}\cdot ||\widehat{A}-A||_{op}\cdot ||(\lambda I + A^* A)^{-1}\widetilde{\varphi}|| \nonumber \\
    & + ||(\lambda (\lambda I + \widehat{A}^*\widehat{A})^{-1}||_{op}\cdot ||\widehat{A}-A||_{op}\cdot ||A(\lambda I + A^*A)^{-1}\widetilde{\varphi}|| \nonumber \\
    & \leq C ||\widehat{A}^*||_{op}\cdot ||\widehat{A}-A||_{op} \lambda^{\frac{\theta \wedge 2}{2}-1} + ||\widehat{A}-A||_{op} \lambda^{\frac{\theta+1}{2}\wedge 1 - 1}\text{ , }
\end{align}
where the last inequality follows from $||(\lambda (\lambda I + \widehat{A}^*\widehat{A})^{-1}||_{op}\leq 2$ (see Lemma \ref{Lemm: Bounds on Compact Operator}(b)),  $||(\lambda I + A^*A)^{-1}\widetilde{\varphi}||\leq C \lambda^{\frac{\theta \wedge 2}{2}-1}$ (see Lemma \ref{Lemm: Bounds on Compact Operator}(d) and Assumption 3.5(b)), and $||A(\lambda I + A^*A)^{-1}\widetilde{\varphi}||\leq C \lambda^{\frac{\theta+1}{2}\wedge 1 - 1}$ (see Lemma \ref{Lemm: Bounds on Compact Operator}(e) and Assumption 3.5(b)). \\
Similarly, the decomposition in (\ref{eq: decomposition for xi4}) leads to
\begin{align}\label{eq: Ahat Xi4}
    ||\widehat{A}\Xi_4||\leq & || \widehat{A} (\lambda I + \widehat{A}^*\widehat{A})^{-1}\widehat{A}^*||_{op}\cdot ||\widehat{A}-A||_{op}\cdot \lambda \cdot ||(\lambda I + A^*A)^{-1}\widetilde{\varphi}|| \nonumber \\
    & + \lambda \cdot ||\widehat{A}(\lambda I + \widehat{A}^*\widehat{A})^{-1}||_{op}\cdot ||\widehat{A}^*-A^*||_{op}\cdot ||A(\lambda I + A^*A)^{-1}\widetilde{\varphi}|| \nonumber \\
    & \leq ||\widehat{A}-A||_{op}\cdot \lambda^{\frac{\theta\wedge 2}{2}}+||\widehat{A}-A||_{op} \cdot \lambda^{\frac{\theta+1}{2}\wedge 1 - 1/2} \text{, }
\end{align}
where the last equality follows from Lemma \ref{Lemm: Bounds on Compact Operator}(a)(c)(d)(e) and Assumption 3.5(b). So, recalling that $\theta\geq 2$, that $\|\widehat{A}-A\|_{op}/\sqrt{\lambda}=o_P(1)$, and  putting together (\ref{eq: Ahat Xi4}), (\ref{eq: bound for Xi4}), and (\ref{eq: inner product of xi4}) gives 
\begin{equation}\label{eq: negligibility of inner product for Xi4}
    \sqrt{n}\left| \left< \Xi_4 , g \right> \right|=O_P\left( \sqrt{n}||\widehat{A}-A||_{op}^2 +\sqrt{n}\lambda\right)=o_P(\sqrt{n}\lambda )=o_P(1)\text{ , }
\end{equation}
where in the last equality we have used $n\lambda^2=o(1)$ (see Assumption 3.4). \\
By gathering (\ref{eq: negligibility of inner product for Xi4}), (\ref{eq: _Xi3}), (\ref{eq: inner product of Regularization Bias }), (\ref{eq: Xi2}), and the decomposition in Equation (\ref{eq: decomposition of phi.hat}), we obtain 

\begin{equation}\label{eq: inner product of phi.hat}
\sqrt{n}\left< \widehat{\varphi}-\widetilde{\varphi}, g \right>=\sqrt{n}\left< \Xi_1 , g \right> + o_P(1) \text{ . }
\end{equation}
So, to show the desired result it suffices to obtain an Influence Function Representation for the leading term of the above display. Now, by the definition of $\Xi_1$ we get 
\begin{align}\label{eq: inner product of Xi1}
    \sqrt{n}\left<  \Xi_1 , g \right> =& \left< \sqrt{n} A^*(\widehat{r}-\widehat{A}\widetilde{\varphi}) ,(A^*A)^{-1}g \right> \nonumber  \\
    & + \left< \sqrt{n} A^*(\widehat{r}-\widehat{A}\widetilde{\varphi}) ,\left[(\lambda I + A^*A)^{-1}-(A^*A)^{-1}\right]g \right>\text{ . }
\end{align}
By using a change of variable, Assumption 3.2 about $\widetilde{\varphi}$, and the $\rho$th order of the kernel $K_Z$, we obtain
\begin{align*}
\widehat{r}(w)-(\widehat{A}\widetilde{\varphi})(w)=& \frac{1}{n h}\sum_{i=1}^n Y_i \, K_W \left(\frac{w-W_i}{h}\right)\frac{1}{\tau(w)}\\
& - \frac{1}{n h}\sum_{i=1}^n  K_W \left(\frac{w-W_i}{h}\right) \frac{ h^{-1}}{\tau(w)} \int_{ }\widetilde{\varphi}(z)\, K_Z \left( \frac{z-Z_i}{h} \right)\, d\, z\\
=& \frac{1}{n h}\sum_{i=1}^n U_i\, K_W \left(\frac{w-W_i}{h}\right) \frac{1}{\tau(w)} \\
& - \frac{1}{n h}\sum_{i=1}^n  K_W \left(\frac{w-W_i}{h}\right) \frac{1}{\tau(w)} \int_{ } \left[\widetilde{\varphi}(Z_i+v h) - \widetilde{\varphi}(Z_i) \right] K_Z \left(v \right)\, d\, v  \\
=& \frac{1}{n h^1}\sum_{i=1}^n U_i K_W \left(\frac{w-W_i}{h}\right)\frac{1}{\tau(w)} \\
& - \frac{h^\rho }{n h^1}\sum_{i=1}^n  K_W \left(\frac{w-W_i}{h}\right)\frac{1}{\tau(w)}  S_{n}(Z_i)\text{ with }|S_{n}(z)|\leq C\text{ , }
\end{align*}
By the previous display and the definition of $A^*$ we obtain (see the comments below) 
\begin{align}\label{decomposition of Astar rhat - Ahat phi0}
    \sqrt{n}\, \left[A^*(\widehat{r}-\widehat{A}\widetilde{\varphi})\right](z)=& \frac{1}{\sqrt{n}}\sum_{i=1}^n \widetilde{U}_i h^{-1}\, \int_{ }\frac{f_{Z W}(z,w)}{\pi(z)\,\tau(w)}\, K_W\left(\frac{w-W_i}{h}\right)\, d\, w \nonumber\\
    &+ \frac{h^\rho}{\sqrt{n}}\sum_{i=1}^n S_n(Z_i)\,h^{-1}\, \int_{ }\frac{f_{Z W}(z,w)}{\pi(z)\, \tau(w)}\, K_W\left(\frac{w-W_i}{h}\right)\,d\,w \nonumber \\
    =& \frac{1}{\sqrt{n}}\sum_{i=1}^n \widetilde U_i \, \int_{ }\frac{f_{Z W}(z,W_i+v h)}{\pi(z)\,\tau(W_i+v h)}\, K_W\left(v\right)\, d\, v \nonumber\\
    &+ \frac{h^\rho}{\sqrt{n}}\sum_{i=1}^n S_n(Z_i)\, \int_{ }\frac{f_{Z W}(z,W_i+v h)}{\pi(z)\, \tau(W_i + v h)}\, K_W\left(v\right)\,d\,w \nonumber \\
    =& \frac{1}{\sqrt{n}}\sum_{i=1}^n \widetilde U_i \frac{f_{Z W}(z,W_i)}{\pi(z)\,\tau(W_i)}+\frac{h^\rho}{\sqrt{n}}\sum_{i=1}^n U_i S_n^{(2)}(W_i,z) \nonumber\\
    &- \frac{h^{\rho}}{\sqrt{n}}\sum_{i=1}^n S_n(Z_i) \frac{f_{Z W}(z,W_i)}{\pi(z)\,\tau(W_i)} - \frac{h^{2\rho}}{\sqrt{n}}\sum_{i=1}^n S_n(Z_i)\,S_n^{(3)}(W_i,z) \\
    &\text{ with }\left|S_n^{(2)}(w,z)\right|\leq C\text{ and } \left|S_n^{(3)}(w,z)\right|\leq C\, , \nonumber
\end{align}
where in the second equality we have used a change of variable, while in the third equality we have combined Assumption 3.2 with the $\rho$th order of the kernel $K_W$. By the iid assumption and since $\mathbb{E}[\widetilde U|W]=0$,

\begin{align*}
    \mathbb{E}\|\text{1st term RHS of (\ref{decomposition of Astar rhat - Ahat phi0})}\|^2 = & \mathbb{E} \frac{1}{n}\sum_{i,j} \widetilde U_i \widetilde U_j\,\int_{ }\frac{f_{Z W}(z,W_i)}{\pi(z)\,\tau(W_i)}\,\frac{f_{Z W}(z,W_j)}{\pi(z)\,\tau(W_j)}\,\pi(d\,z) \\
    =& \int_{ }\,\left[\frac{f_{Z W}(z,w)}{\pi(z)\,\tau(w)}\right]^2\frac{\sigma_W(w)^2\,f_W(w)}{\tau(w)}\,\,\pi(d z)\otimes\tau(d w)\,, 
\end{align*}
where $\sigma_W^2:=\mathbb{E}[\widetilde U^2|W=\cdot]$. Assumption 3.2 ensures that $\sigma_W^2 f_W/\tau$ is bounded and $f_{Z W}/(\pi\,\tau)\in L^2_{\pi\otimes\tau}$, so the RHS of the above display is finite. Similarly,  
\begin{align*}
    \mathbb{E}\|\text{2nd term RHS of (\ref{decomposition of Astar rhat - Ahat phi0})}\|^2 = &h^{2\rho} \int_{ }\frac{\sigma_W^2(w) f_W(w)}{\tau(w)} \left|S_n^{(2)}(w,z)\right|^2 \pi(dz)\otimes\tau(dw)\\
    =& O(h^{2\rho})=o(1)\, ,
\end{align*}
where the last equality has used the fact that $|S_n^{(2)}(w,z)|\leq C$. Similar arguments show that 
\begin{align*}
    \mathbb{E}\|\text{3rd term RHS of (\ref{decomposition of Astar rhat - Ahat phi0})}\|^2 =& \frac{h^{2 \rho}}{n} \sum_{i,j} \mathbb{E}[S_n(Z_i) S_n(Z_j)]\int_{ }\frac{f_{Z W}(z,W_i)}{\pi(z)\,\tau(W_i)} \frac{f_{Z W}(z,W_j)}{\pi(z)\,\tau(W_j)}\, \pi(d z)\\
    \leq & C\,h^{2 \rho} \int_{ }\left[\frac{f_{Z W}(z,w)}{\pi(z)\,\tau(w)}\right]^2 \frac{f_W(w)}{\tau(w)} \, \pi(d z)\otimes \tau(d w) \\
    &+ C\, h^{2 \rho} \frac{n(n-1)}{n} \int_{ }\left\{\int_{ }\frac{f_{Z W}(z,w)}{\pi(z)\,\tau(w)}\frac{f_W(w)}{\tau(w)}\,\tau(d w)\right\}^2 \pi(d z) \\
    \leq & C h^{2\rho} n \int_{ } \left[\frac{f_{Z W}(z,w)}{\pi(z)\,\tau(w)}\right]^2 \pi(d z)\otimes \tau(d w) \\
    =& o(1)\, ,
\end{align*}
where for the first inequality we have used the boundedness of $S_n$, for the second inequality we have used the boundedness of $f_W/\tau$ (see Assumption 3.2)  and the Cauchy-Schwartz inequality, while for the last equality we have used Assumption 3.4. \\
By proceeding along the same lines we find that 
\begin{equation*}
    \mathbb{E}\|\text{4th term RHS of (\ref{decomposition of Astar rhat - Ahat phi0})}\|^2 \leq C h^{4 \rho} n=o(1)\, .
\end{equation*}

We are now able to obtain the Influence function representation for $\sqrt{n}\left<\Xi_1,g\right>$ in Equation (\ref{eq: inner product of Xi1}). By the previous five displays and since $\|(A^*A)^{-1}g\|<\infty$, the first term on the RHS of (\ref{eq: inner product of Xi1}) equals
\begin{align*}
    \left<\sqrt{n}A^*(\widehat r - \widehat A \widetilde{\varphi}) , (A^*A)^{-1}g\right>=&\left<\frac{1}{\sqrt{n}}\sum_{i=1}^n \widetilde U_i \frac{f_{Z W}(\cdot,W_i)}{\pi(\cdot)\tau(W_i)}, (A^*A)^{-1}g\right>+o_P(1)\\
    = & \frac{1}{\sqrt{n}}\sum_{i=1}^n \widetilde  U_i \int_{ }\frac{f_{Z W}(z,W_i)}{\tau(W_i)}\left[(A^*A)^{-1}g\right](z)\, d z +o_P(1)\\
    =& \frac{1}{\sqrt{n}}\sum_{i=1}^n \widetilde U_i \left[A(A^*A)^{-1}g\right](W_i)+o_P(1)\, .
\end{align*}
To conclude the proof, notice first that Equation (\ref{decomposition of Astar rhat - Ahat phi0}) and the four displays after it ensure that $\|\sqrt{n}A^*(\widehat r - \widehat A \widetilde{\varphi})\|=O_P(1)$.  So, the norm of the second term on the RHS of (\ref{eq: inner product of Xi1}) is upperbounded as follows  
\begin{align*}
    \left|\left<\sqrt{n}A^*(\widehat r - \widehat A \widetilde{\varphi}),\left[(\lambda I + A^* A)^{-1}-(A^*A)^{-1}\right]g\right>\right|\leq O_P\left( \|\left[(\lambda I + A^* A)^{-1}-(A^*A)^{-1}\right]g\| \right) \, .
\end{align*}
By recalling that we have set $\eta=2$, we have $(A^*A)^{\eta/2}=A^*A$. Thus, by definition of $\widetilde{f}$ we get $g=(A^*A)^{\eta/2}\widetilde{f}=A^*A\widetilde{f}$ and 
\begin{align*}
    \left[(\lambda I + A^* A)^{-1}-(A^*A)^{-1}\right]g=(\lambda I + A^* A)^{-1}A^*(A\widetilde{f})-\widetilde{f}\, .
\end{align*}
Since $(\lambda I + A^*A)^{-1}A^*$ is a regularization scheme, the RHS of the above display converges to 0 as $\lambda\rightarrow 0$ (see \citet{kress_linear_1999}). This concludes the proof.

\end{Proof}

\subsubsection{Auxiliary Results}

The following lemma is borrowed from \citet{florens_instrumental_2012} (see their Lemma A.1(b)).
\begin{lemma}\label{Lem: rates for Ahat and rhat}
Under Assumptions 3.2 to 3.4 in the main text, we have
$$||\widehat{A}-A||_{op}=O_P\left( \frac{1}{\sqrt{nh}}+ h^\rho \right) \text{ and }||\widehat{r}-r||=O_P\left( \frac{1}{\sqrt{nh}}+h^\rho \right)$$ 
\end{lemma}
The following lemma is well established in the literature on empirical process theory. A proof can be found in \citet{andrews_chapter_1994}. 

\begin{lemma}\label{ASE lemma}
Let $T$ be a random variable such that $\mathbb{E}[T|Z]$ is bounded and let $\mathcal{F}$ be a class of functions of $Z$ such that $\int_{0}^1\sqrt{N_{[\,\, ]}(\epsilon,\mathcal{F},\|\cdot \|_{\P})}\, d\, \epsilon<\infty$. If   $\|(\widehat{f}-f_0)T\|_{2,P}=o_P(1)$ and $\P(\widehat{f}\in\mathcal{F})\rightarrow 1$, then 
$$\sqrt{n}\, \left( \mathbb{P}_n-\P \right)\, \left( \widehat{f}-f_0 \right)T=o_P(1)\, .$$
\end{lemma}

The lemma that follows gathers several useful results about inequalities of norms involving compact operators. Its proof can be found in \citet{florens_identification_2011} and \citet{carrasco_chapter_2007}.

\begin{lemma}\label{Lemm: Bounds on Compact Operator}
Let $\mathcal{X}$ and $\mathcal{Y}$ be two Hilbert spaces and $K:\mathcal{X}\mapsto \mathcal{Y} $ be a linear compact operator with Singular Value Decomposition given by $(\widetilde{\lambda}_j,\widetilde{\varphi}_j,\widetilde{\psi}_j)_j$. \\
\begin{itemize}
\item[(a)] $$||K(\lambda I + K^*K)^{-1}K^*||_{op}\leq 1 $$
\item[(b)] $$||\lambda(\lambda I + K^* K)^{-1}||_{op}\leq 2 $$
\item[(c)] $$||(\lambda I + K^* K )^{-1}K^* ||_{op}\leq \frac{1}{2\sqrt{\lambda}}$$

    \item[(d)]  If $||\phi||_\gamma^2:=\sum_j\widetilde{\mu}_j^{-2\gamma}|\left<\phi,\widetilde{\varphi}_j\right> |^2<\infty$ then 
    $$||\lambda(\lambda I + K^*K)^{-1}\phi||\leq C ||\phi||_\gamma \lambda^{\frac{\gamma\wedge 2}{2}}$$
    \item[(e)] Under the same conditions of (d) it holds that  $$||\lambda K (\lambda I + K^*K)^{-1} \phi ||\leq C||\phi||_\gamma \lambda^{\frac{\gamma+1}{2}\wedge 1 }  $$
    \item[(f)] If $||\phi||^2_\gamma<\infty$ then 
    $$||(\lambda I + K^*K)^{-1}K^*K\phi - \phi ||=O(\lambda^{\frac{\gamma \wedge 2}{2}})$$

\end{itemize} 
\end{lemma}

\begin{lemma}\label{Lem: rate for phihat}
Under Assumptions 3.1 to 3.6 in the main text, we have
$$||\widehat{\varphi}-\widetilde{\varphi}||_{\pi}=O_P\left( \frac{||\widehat{A}-A||}{\sqrt{\lambda}}+\frac{||\widehat{r}-r||}{\sqrt{\lambda}}+\lambda^{\frac{\theta \wedge 2}{2}}  \right)$$
where $||\cdot||$ denotes the $L^2(\mathbb{R})$ norm
\end{lemma}
\begin{Proof}
The proof readily follows from the decomposition in (\ref{eq: decomposition of phi.hat}) and Lemma \ref{Lemm: Bounds on Compact Operator}. 
\end{Proof}
\subsection{Power analysis in the nonparametric framework}\label{app.C}
We consider the scalar case $p=q=1$. It is clear that our test will have asymptotic power equal to one against alternatives for which $\E[\widetilde U (W-\E[W])X]\ne 0$. Remark that $\E[\widetilde U (W-\E[W])X]$ can take in principle any value. So, we argue that the alternatives against which we do not have power are degenerate. Let us give some examples.\\

\noindent \textbf{Example 3.}
Let $E$ be uniformly distributed on the interval $[-1,1]$. Let also $X$ and $W$ be Bernoulli random variables with parameters equal to $1/2$, respectively. The variables $E, X, W$ are mutually independent. The treatment is generated as 
$$Z=WXI(E\ge 0).$$
The residual $U(0)$ and $U(1)$ follow
\begin{align*}
    U(0) &=\rho XE;\\
    U(1)&=0,
\end{align*}
where $\rho>0$.
We assume that $\varphi(0)=\varphi(1)=0$, so that $Y=U(Z).$ Notice that $Y=0$ when $Z=0$ and $Y=\rho XE$ when $Z=1$. The mapping $\widetilde \varphi$ solves
\begin{equation} \label{solphi}\E[Y|W]=\E[\widetilde \varphi (Z)|W].\end{equation}
Using the fact that, on the event $\{W=0\}$, we have $Z=0$, we get 
\begin{align*}\widetilde \varphi(0)=\E[Y|W=0]=0.
\end{align*}
Next, we have 
\begin{align}
\label{esp}\E[Y|W=1]&= \frac14   \E[Y|E\ge 0, X=1,W=1]=\frac\rho8  .
\end{align}
Since $\widetilde \varphi(0)=0$ and $\P(Z=1|W=1)=\frac14$, it also holds that $\E[\widetilde \varphi (Z)|W=1] =\frac14\widetilde\varphi(1)$. By \eqref{esp}, this leads to $\widetilde\varphi(1)=-\frac{\rho}{2}$. As a result,  we obtain $\widetilde U =Y-\widetilde \varphi(Z)=Z(\rho XE-\frac{\rho}{2}).$
Hence, we have 
\begin{align*} \E[\widetilde UWX]&=\frac{1}{4} \E[\widetilde U|W=1, X=1]\\
&=\frac{1}{4} \E\left[\left.Z\left(\rho XE-\frac{\rho}{2}\right)\right|W=1, X=1\right]\\
&=\frac{\rho}{8} (\E[E|E\ge 0, W=1, X=1]-1)=-\frac{\rho}{16}.
\end{align*}
Therefore, in this example, our test does not have power only when $\rho =0$, which is a degenerate case.\\

\noindent \textbf{Example 4.} Let $(W,E,X)^\top$ be a $3\times 1$ Gaussian vector with mean zero and variance equal to the identity matrix. Let also $Z=W+E+X$, $U(Z) = Z(E+\rho X)$, where $\rho \in\R$, and $\varphi \equiv 0$ so that $Y=U(Z)=Z(E+\rho X)$. In this case, we have 
\begin{align*}
   \E[Y|W]= \E[Z(E+\rho X)|W]=\E[WE+E^2+XE+\rho(WX+EX+X^2)|W]=1+\rho.
\end{align*}
Hence, $\widetilde \varphi \equiv 1+\rho $ solves the equation. It is the unique solution since $Z$ is strongly complete conditional on $W$ (see \cite{newey2003instrumental} for a discussion of conditional completeness in the Gaussian case). As a result, we have $\widetilde{U} = Z(E+\rho X)-1-\rho$. Hence, we get 
\begin{align*}
    \E[\widetilde U WX]&= \E[(W+E+X)(E+\rho X)WX]-(1+\rho)\E[WX]\\
    &=\rho \E[(WX)^2]=\rho.
\end{align*}
As a result, the test does not have power only when $X$ is not correlated with $U$, i.e. $\rho =0$, which is a degenerate case.

\bibliographystyle{chicago}
\bibliography{supplement}
